\documentclass[10pt,journal]{IEEEtran}

\usepackage{graphics}
\usepackage{color,soul}
\usepackage{graphicx}
\usepackage{times}
\usepackage[numbers,sort&compress]{natbib}
\usepackage{hyperref}
\usepackage{subcaption}
\usepackage{fancyhdr}
\usepackage{booktabs}
\usepackage{multirow}
\usepackage{amssymb}
\usepackage{hyperref}
\usepackage{color,soul}
\usepackage{tabularx}
\usepackage{enumitem}
\usepackage{float}
\usepackage{multirow}
\usepackage{colortbl}
\usepackage[table]{xcolor}
\usepackage{makecell}
\usepackage{pdflscape}
\usepackage{nicematrix}
\usepackage{color}
\usepackage{graphicx}
\usepackage{listings}
\usepackage{floatflt}
\usepackage{xcolor}
\usepackage{graphicx}
\usepackage{array}
\usepackage{caption}
\usepackage{lipsum}
\usepackage{booktabs}
\usepackage{fancyhdr}

\newcommand{\takeaway}[1]{\textit{\textbf{Takeaway:}} \textit{#1}}

\usepackage{mdframed}
\definecolor{gray90}{gray}{0.90}
\definecolor{gray95}{gray}{0.95}
\newenvironment{summary}
{\vspace{-1em}\noindent\begin{mdframed}[backgroundcolor = gray90, linecolor = black, innerleftmargin=2mm, innerrightmargin=2mm]}
{\end{mdframed}}

\fancypagestyle{FirstPage}{
}

\begin{document}
%\title{Evaluating DNS Resiliency and Response Times with Truncation, Fragmentation \& DoTCP Fallback}
\title{Evaluating DNS Resiliency and Responsiveness with Truncation, Fragmentation \& DoTCP Fallback}

\author{\IEEEauthorblockN{Pratyush Dikshit\IEEEauthorrefmark{1},
Mike Kosek\IEEEauthorrefmark{2},
Nils Faulhaber\IEEEauthorrefmark{2}, 
%Simone Ferlin\IEEEauthorrefmark{3}, 
Jayasree Sengupta\IEEEauthorrefmark{1}, and
Vaibhav Bajpai\IEEEauthorrefmark{1}}\\
\IEEEauthorblockA{\IEEEauthorrefmark{1}CISPA Helmholtz Center for Information Security, Germany}\\
\IEEEauthorblockA{\texttt{[pratyush.dikshit $|$ jayasree.sengupta $|$ bajpai]@cispa.de}}\\
\IEEEauthorblockA{\IEEEauthorrefmark{2}Technical University of Munich, Germany}\\
\IEEEauthorblockA{\texttt{[kosek $|$ nils.faulhaber]@tum.de}}\\
%\IEEEauthorblockA{\IEEEauthorrefmark{3}Red Hat, Sweden}% <-this % stops an unwanted space
%\IEEEauthorblockA{\texttt{[simone@ferlin.io]}}
}

\maketitle
\IEEEpeerreviewmaketitle

\begin{abstract}

Since its introduction in 1987, the DNS has become one of the core components of the Internet. While it was designed to work with both TCP and UDP, DNS-over-UDP (DoUDP) has become the default option due to its low overhead. As new Resource Records were introduced, the sizes of DNS responses increased considerably. This expansion of message body has led to truncation and IP fragmentation more often in recent years where large UDP responses make DNS an easy vector for amplifying denial-of-service attacks which can reduce the resiliency of DNS services. This paper investigates the resiliency, responsiveness, and usage of DoTCP and DoUDP over IPv4 and IPv6 for 10 widely used public DNS resolvers. In these experiments, these aspects are investigated from the edge and from the core of the Internet to represent the communication of the resolvers with DNS clients and authoritative name servers. Overall, more than 14M individual measurements from 2527 RIPE Atlas Probes have been analyzed, highlighting that most resolvers show similar resiliency for both DoTCP and DoUDP. While DNS Flag Day 2020 recommended 1232 bytes of buffer sizes yet we find out that 3 out of 10 resolvers mainly announce very large EDNS(0) buffer sizes both from the edge as well as from the core, which potentially causes fragmentation. In reaction to large response sizes from authoritative name servers, we find that resolvers do not fall back to the usage of DoTCP in many cases, bearing the risk of fragmented responses. As the message sizes in the DNS are expected to grow further, this problem will become more urgent in the future.

\end{abstract}


\section{Introduction}
\label{introduction}

\begin{figure}[!t]
    \centering
    \includegraphics[width=1\linewidth]{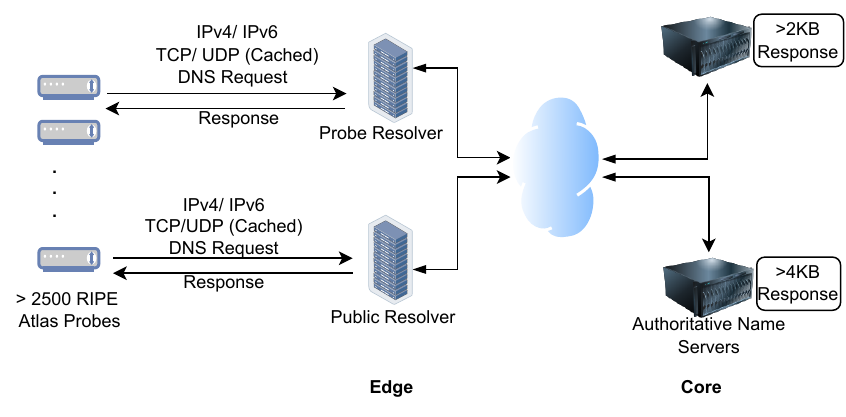}
    \caption{\em 2527 RIPE Atlas Probes communicate the DNS requests with the edge (Probe and Public Resolvers) and with the core (authoritative NSes) using IPv4 and IPv6. Cached DNS responses are sent by the edge, while uncached DNS responses (2KB and 4KB) are sent by the core.}
    \label{fig:Intro}
    \vspace{-1em}
\end{figure}

\thispagestyle{FirstPage}

The Domain Name System (DNS), which is responsible for the resolution of hostnames to IP addresses, has become one of the most widely used components on the Internet. Hostnames (domain names) are organized in a tree structure that is hierarchically separated into \emph{zones}. The resolution of domain names is realized by different components such as stub resolvers, recursive resolvers, and authoritative Name Servers (NSes). While authoritative NSes are responsible for the authoritative mapping of domains in a zone to their IP addresses, stub, and recursive resolvers cache and deliver such information from the NSes to the clients via a DNS request \cite{DNSDevelopment_Paul} (RFC 1034 \cite{IETF_rfc1034}). DNS communication supports both major transport protocols on the Internet, namely the Transmission Control Protocol (TCP) (RFC 793 \cite{IETF_rfc793}) and the User Datagram Protocol (UDP) (RFC 768 \cite{IETF_rfc768}). Due to its comparably low overhead, UDP has become the default transport protocol for DNS. The UDP message body is restricted to 512 bytes (RFC 1035 \cite{ietf_RFC_1035}). However, the increase in deployment of DNS Security (DNSSEC) and IPv6 (RFC 7766 \cite{IETF_rfc7766}) has resulted in larger message sizes, thereby leading to two important developments in the protocol. Firstly, DNS-over-TCP (DoTCP) was declared to be mandatory for hosts (RFC 5966 \cite{IETF_rfc5966}) as it enables a larger message body by default. Secondly, Extension Mechanisms for DNS (EDNS) were introduced to augment the capabilities of the DNS protocol in terms of message size expansion (RFC 2671 \cite{IETF_rfc2671}). With the new EDNS capability, it was required that DNS replies would continue to provide responses as UDP datagrams even though the response was larger than 512 bytes. Stipovic \emph{et al.} in \cite{Ivica_DNS_Compliance} examines the level of compatibility for a number of public DNS servers for some popular Internet domains while exploring the behavior of some contemporary DNS implementations such as Microsoft Windows 2012, 2016 and 2019 as well as Linux-based BIND in regards to the EDNS. However, using too large UDP buffer sizes can cause IP fragmentation in certain networks, thereby reducing resiliency in DNS communication \cite{koolhaas_defragmentingDNS}. To avoid fragmentation, the DNS Flag Day, 2020\footnote[1]{\label{DNS Flag Day}\url{http://www.dnsflagday.net/2020/}}, an association of DNS software maintainers and service providers, recommended the usage of a default buffer size of 1232 bytes. DoTCP is a useful measure against fragmentation and can increase DNS resiliency by allowing fallback options. Resolvers should also avoid fragmentation by using the recommended default EDNS(0) buffer size of 1232 bytes. To this end, our paper puts forward three \textbf{goals:} \emph{a)} to evaluate DoTCP support (both over IPv4 and IPv6) and its usage across several DNS resolvers, \emph{b)} to analyze the responsiveness/ latency over DoTCP and DoUDP for IPv4 and IPv6, and \emph{c)} to investigate which buffer sizes are currently used in DNS traffic around the globe.

In pursuit of these goals, we evaluate the behavior of the resolvers from two different vantage points. Firstly, DoTCP adoption, responsiveness, and EDNS(0) configuration are analyzed from the edge where the interaction between recursive resolvers and DNS clients running on the RIPE Atlas probes is measured. To scope DNS requests to the \emph{Edge} of the network, we perform DNS queries for a domain that is likely cached by all resolvers, unlike in previous studies \cite{DoTCP_Kosek}. Secondly, the interaction of recursive resolvers with authoritative NSes is further studied. To allow DNS requests to leave the edge and move into the \emph{Core} of the network, we provision dedicated NSes for a custom-crafted domain whose resolution is requested from the DNS resolvers. Using this methodology (see §\ref{methodology}), we study failure rates, response times, EDNS options, and usage of DoTCP and DoUDP, as well as the EDNS(0) configurations both from the edge and the core (except response times analysis) that gives detailed insights into the potential resiliency of DNS communication on the Internet as depicted in Figure \ref{fig:Intro}. We perform measurements over IPv4 and IPv6 \cite{bajpai:ton:2019}. Our main \textbf{findings} (see §\ref{results}) are -- 
%\subsubsection*{RIPE Atlas Probes} All the RA probes (2527) are %IPv4 capable and 1137 probes are IPv6 capable with the desired %attributes. North America (18\%) and Europe (70\%) have highest %density of probes.
\subsubsection*{\textbf{Resiliency from the edge}} We observe that DoTCP (4.01\%) tends to fail less often than DoUDP (6.3\%) requests over IPv4. Contrarily, in the case of IPv6, we find a higher failure rate over both transport protocols (DoTCP 10\%, DoUDP 9.61\%). The analysis of response times for Public and Probe resolvers confirms the pattern of approximately doubled median response times for DoTCP compared to DoUDP in both the IP versions. We also observe that several public DNS resolvers still lack adoption ($<3.5\%$) of 1232B from the DNS Flag Day recommendation.
\subsubsection*{\textbf{Resiliency from the Core}} We find that DoTCP requests over IPv4 exhibit failure rates of 9.09\% on public resolvers against higher failure rate of 11.53\% over IPv6. Surprisingly, we find that the RIPE Atlas measurements ended successfully even after receiving a response with the TC-bit set, indicating a lack of proper fallback to DoTCP in many probes. Moreover, communication between resolvers and the authoritative NSes utilizes an EDNS(0) buffer size of 512 bytes less preferably (IPv4 0.24\%, IPv6 0.13\%) compared to the buffer sizes advertised to the RIPE Atlas probes (IPv4 27.41\%, IPv6 26.04\%). All DNS resolvers use EDNS(0) in most of the cases ($>$ 99.84\%). We also see other DNS options such as Cookie (4.80\% IPv4, 7.91\% IPv6) and EDNS Client Subnet (ECS) (1.81\% IPv4, 1.49\% IPv6) advertised by the public resolvers, while Google mostly uses ECS (14.24\% IPv4, 12.53\% IPv6).
\subsubsection*{\textbf{DoTCP Usage Rates}} We observe that when 2KB responses are received from the NSes, all resolvers that mainly use canonical (see §\ref{results} scenarios, use TCP in their last request for $>$95\% of the cases. In situations where 4KB responses are received, we observe that almost all resolvers use TCP in the vast majority of measurements over both IP versions ($>$98\%).

This paper builds upon our previous study \cite{Internet_resilience_IFIP23}. In this paper, we have additionally added significant background information (see:~§\ref{Related_Work}) related to the monitoring and performance evaluation of DNS query-response over both TCP and UDP transport protocols. DNS response times can be a critical metric when using DoTCP fallback, we therefore conduct further measurements comparing DoTCP and DoUDP response times from the edge of the network (see: §\ref{edge_response times results}). Subsequently, when evaluating from the core, we present additional insights by including a detailed analysis using \texttt{traceroute} and DoUDP response time for public resolvers (see: §\ref{Traceroute}). %We have also added the details of the distribution of resolvers communicating with authoritative name servers for uncached domains (see §~\ref{table2:Distribution_backend}). 
Additionally, we perform a deep dive by measuring the number of successful responses that do not contain any valid \texttt{ANSWER} sections for the DNS queries (see: §\ref{Valid response}). Notably, our investigation reveals instances where RIPE Atlas measurements have successfully terminated, despite receiving a response with the \texttt{TC}-bit set, thereby, indicating a lack of proper fallback to DoTCP across multiple probes. Towards the end, we discuss the limitations of our study and highlight future research directions in §\ref{limitations}, followed by concluding statements in §\ref{conclusion}.

%into a detailed analysis of the measurement of 

%comparison between these two implementations.

%Since response time represents a primary drawback when employing DoTCP instead of DoUDP, we conduct a thorough comparison between the two implementations to evaluate this aspect in depth.

\section{Background and Related Work}
\label{Related_Work}

% \usepackage{booktabs}
% \usepackage{graphicx}
\begin{table}[]
\caption{\em The public DNS resolvers under investigation, IPv4 and IPv6 represent the resolvers’ anycast addresses}
\centering
\resizebox{\columnwidth}{!}{%
\begin{tabular}{@{}lll@{}}
\toprule

\textbf{Name}           & \textbf{IPv4}   & \textbf{IPv6}        \\ \midrule
 
Cloudlfare Public DNS   & 1.1.1.1         & 2606:4700:4700::1111 \\
 
Google Public DNS       & 8.8.8.8         & 2001:4860:4860::8888 \\
 
CleanBrowsing           & 185.228.168.9   & 2a0d:2a00:1::        \\

OpenDNS                 & 208.67.222.222  & 2620:119:35::35      \\
 
OpenNIC                 & 185.121.177.177 & 2a05:dfc7:5::53      \\
 
Quad9                   & 9.9.9.9         & 2620:fe::fe          \\

Comodo Secure DNS       & 8.26.56.26      & -                    \\
 
UncensoredDNS           & 91.239.100.100  & 2001:67c:28a4::      \\
 
Neustar UltraDNS Public & 64.6.64.6       & 2620:74:1b::1:1      \\
 
Yandex.DNS              & 77.88.8.8       & 2a02:6b8::feed:ff    \\ \bottomrule
\end{tabular}%
}

\label{Public_DNS_resolvers}
\end{table}

\subsection{DNS Measurement}
To measure DNS failure rates, DNS performance, and the buffer sizes used, several studies have been conducted in the last few years. Some of them are discussed here.

\subsubsection{\textbf{Fragmentation}}
\label{Fragmentation}

With the increased message sizes, DNS queries can exceed the MTU of many networks. Giovane \emph{et al.} in \cite{fragmentation_moura} analyzes the fragmentation rates of DNS queries to the \textit{.nl} top-level domain showing that less than 10k of 2.2B observed DNS responses by authoritative NSes are fragmented. Although fragmentation is in general fairly rare in DNS communication, the consequences can have negative effects on the resiliency and connectivity of Internet applications (RFC 8900 \cite{IETF_rfc8900}). Herzberg and Shulman in \cite{Fragmentation_Amir} presented an attack allowing to spoof of Resource Records (RRs) in fragmented DNS responses by which attackers can hijack domains or nameservers under certain conditions. Following a similar procedure, Shulman and Waidner in \cite{Fragmentation_leaking_Shulman} showed the opportunity to predict the source port used by the client. Both of these approaches belong to the class of DNS cache poisoning attacks, one of the most common and dangerous on the DNS. Though, cache poisoning attacks are also possible when DNS messages are not fragmented \cite{DNSPoisoning_Sooel}\cite{DNS_Poisoning_BergerDG21}\cite{CERT_DNS_poisoning}. The aforementioned studies however show the additional security risk caused by fragmented responses. This potentially exposes the DNS user to several other types of attacks. Koolhaas \emph{et al.} in \cite{koolhaas_defragmentingDNS} analyzed the behavior of different EDNS(0) buffer sizes in 2020. It was shown that the likelihood of a failing DNS query increases with growing buffer sizes. For a size of 1500 bytes, the default MTU of Ethernet which causes fragmentation of most of the DNS messages of the DNS queries to stub resolvers failed for 18.92\%  over IPv4, with 26.16\% over IPv6 (RFC 2464 \cite{IETF_rfc2464}). As countermeasures, in 2017, Cao \emph{et al.} in \cite{Encoding_DNS_CaoMWL17} presented an ”Encoding scheme against DNS Cache Poisoning Attacks”. Berger \emph{et al.} in \cite{DNS_Poisoning_BergerDG21} presented a way of detecting DNS cache poisoning attacks in 2019. Even though it was later shown that DNS cache poisoning attacks are also possible against DoTCP by Dai \emph{et al.} in \cite{DoTCP_Poisoning_Dai}, this emphasizes its importance as a fallback option to the usage of DoUDP. Herzberg and Shulman in \cite{Fragmentation_leaking_Shulman} recommend keeping the indicated buffer size less or equal to 1500 bytes. As a consequence, Weaver \emph{et al.} summarize a list of recommendations to stakeholders in the DNS ecosystem in \cite{weaver_dnsMeasurements}. These include the proposition for stub resolvers as well as authoritative nameservers to stick to buffer sizes of 1400B and below. The study conducted in \cite{koolhaas_defragmentingDNS} in 2020 yielded detailed recommendations for the EDNS(0) buffer size configuration of authoritative name servers and stub resolvers dependent on the used IP version and network type. The recommendations were adopted at the DNS Flag Day, 2020 claiming that ”defaults in the DNS software should reflect the minimum safe size which is 1232 bytes”. The aforementioned aspects emphasize the need for DNS resolvers to adopt the buffer size recommendations as fast as possible. Some other encrypted DNS protocol implementations such as DNS-over-TLS (RFC 7858 \cite{DoTLS_rfc7858}) and DNS-over-HTTPS (RFC 8484 \cite{DoH_rfc8484}) also counter the problems of fragmentation as TCP used as transport protocol \cite{Dikshit_DNS_Privacy}. They are however not yet widely enough adopted to obsolete standard DNS implementations \cite{DNS_Encryption_Chaoyi}\cite{DNS_privacy_Deccio}\cite{DoTLS_Doan}. To investigate the progress of DNS resolvers in implementing the new standards, measurements analyzing the buffer sizes used by DNS resolvers are therefore performed from different standpoints. As DoTCP support is another important requirement for DNS resolvers to avoid truncation and fragmentation, DNS failure rates over TCP and UDP are analyzed in this paper. Additionally, the DoTCP-fallback behavior of the resolvers is studied to see in which cases they make use of TCP. As furthermore response time is the main disadvantage when using DoTCP instead of DoUDP, we are also interested in comparing the two implementations with regard to this aspect.

\begin{figure}[!t]
    \centering
    \includegraphics[width=1\linewidth,trim={0.2cm 0 3cm 0},clip]{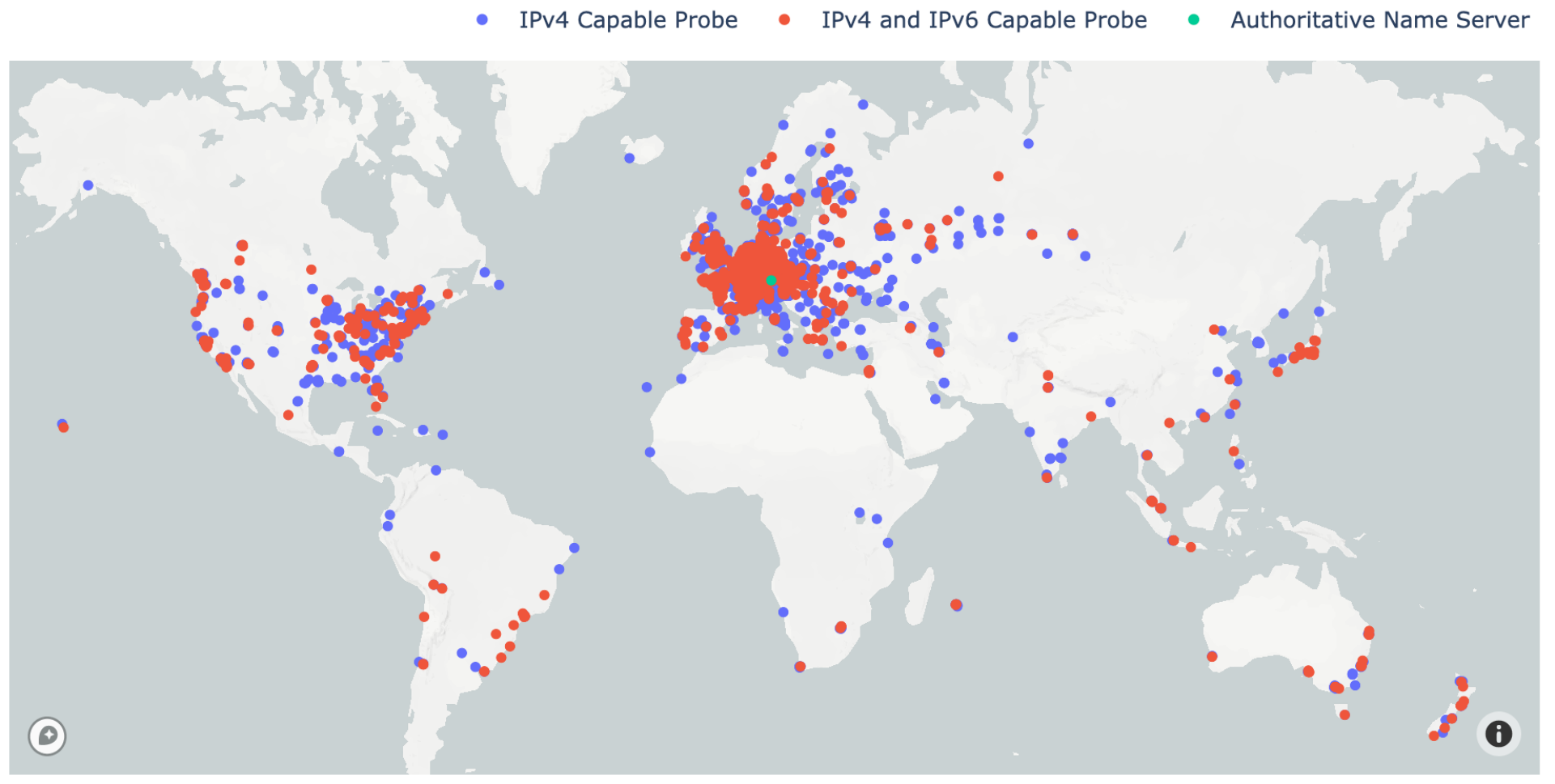}
    \caption{\em The global distribution of the RIPE Atlas probes participating in the experiments and the authoritative name servers developed for the measurements from the core.}
    \label{fig:probe_distribution}
    \vspace{-1em}
\end{figure}

\subsubsection{\textbf{Response Times and Failure Rates}}
\label{Background_Response times}

The first large-scale study on DNS performance and failure rates was performed by Danzig \emph{et al.} in 1992 resulting in several important recommendations to reduce DNS traffic and latency \cite{DNS_Danzig}. Ten years later, DNS Performance and the Effectiveness of Caching analyzed DNS traffic of the MIT Laboratory for Computer Science and the KAIST (Korea Advanced Institute of Science and Technology) over several months and stated a failure rate of 36\% (23\% timeouts, 13\% other errors) as explained by Jung \emph{et al.} in \cite{DNS_Performance_Jung}. Ager \emph{et al.} \cite{Response_time_Ager} compared DNS response times between local DNS resolvers of ISPs and public resolvers like Google, PublicDNS, and OpenDNS. It was found that local resolvers generally outperformed public resolvers, but Google and OpenDNS showed faster responses in certain cases due to ISP caching issues. Several other measurements
have been undertaken to observe DNS performance over IPv4 from different
standpoints \cite{DNS_Evolution_Otto}\cite{Proactive_Caching_Cohen}\cite{DNSPerformance_lookups_Park}. Additionally, Doan \emph{et al.} in \cite{PublicDNS_Doan} observed that the public resolvers answered faster over IPv6 than over IPv4. In 2022, Moura \emph{et al.} \cite{fragmentation_moura} investigated the fallback capabilities of DNS resolvers to utilize DoTCP by manipulating the TC-bit in responses from a controlled authoritative name server. They analyzed the order of incoming requests over different transport protocols and introduced the distinction between canonical (UDP request followed by TCP request) and non-canonical scenarios. Evaluating the order of incoming requests, it was concluded that an estimated 2.7\% (optimistic estimation) to 4.8\% (pessimistic estimation) of the examined resolvers were incapable of falling back to DoTCP usage. In the same year, Kosek \emph{et al.} \cite{DoTCP_Kosek} conducted the first study comparing DNS response times and failure rates based on the underlying transport protocols using RIPE Atlas for ten public resolvers and probe resolvers. 8\% of the queries over UDP and TCP failed, with a very high DoTCP failure rate of 75.0\% for probe resolvers. The response times of DoTCP were generally higher than that of DoUDP with large differences. The queried domains were unique and thereby uncached by all of the participating resolvers. To get a preferably broad and unbiased comparison of the different DNS resolvers over the particular underlying protocols, we perform DNS queries to both the domains, which is very likely cached on each resolver \textit{(google.com)}, and uncached ones. Additionally, we make sure that the uncached domains are administered by an authoritative name server under our control. Using a cached domain allows a detailed estimation of the latencies in the direct communication between resolvers and client software without the additional time needed for a recursive lookup. The DNS queries for uncached domains force recursive resolvers to forward them to our server. Observing the incoming requests offers the opportunity to analyze the communication between recursive resolvers and name servers, for example, the usage of DoTCP on the same path, in detail.

\subsubsection{\textbf{EDNS Options}}

Van den Broek \emph{et al.} \cite{DNSSEC_Fragmentation_Broek_Commag} analyzed more than 8 million DNS queries to an authoritative name server in 2014. Around 75\% of the queries used EDNS(0). Additionally, it was observed that 36\% used a UDP buffer size higher than 1232 bytes likely causing fragmentation. Measurements after the DNS Flag Day 2020 recommendations show that DNS resolvers still seem to lack adoption to the default buffer size of 1232 bytes. Based on the analysis of 164 billion queries to authoritative name servers, Moura \emph{et al.} stated that many resolvers ”announce either small (512 bytes) or large (4096 bytes) EDNS(0) buffer sizes, both leading to more truncation, and increasing the chances of fragmentation/packets being lost on the network”. %Simon Huber produced similar results by comparing the buffer size usage of different public resolvers from the edge using RIPE Atlas. Five out of the ten public resolvers under observation advertised either 512 or 4096 bytes in a vast majority of their responses. Four resolvers mainly conformed to the recommended 1232 bytes [10]. 
As the DNS Flag Day 2020 recommendations have not been out in the community for a very long time, a regular examination of the adoption rates of DNS software is reasonable and necessary. %This paper continues the measurements of Kosek \emph{et al.} \cite{DoTCP_Kosek} to observe changes.

\section{Methodology}
\label{methodology}

To extend the previous studies by Huber and Kosek \emph{et al.} \cite{DoTCP_Kosek}, we utilize the identical target resolvers for our measurements. We query the ten public resolvers listed in Table \ref{Public_DNS_resolvers}, along with the configured probe resolvers. It is worth mentioning that at the time of the initial measurements, Comodo Secure DNS did not have a known IPv6 address, resulting in measurements conducted solely over IPv4.

\subsection{Probe Selection}

This paper employs the RIPE Atlas measurement network to conduct the measurements. To avoid potential load issues occurring in the first two probe versions \cite{bajpaiccr2015}\cite{RIPE_Holterbach}, we choose only probes of version 3 or 4 that are hosted with a hometag \cite{bajpai:im:2017}. The chosen probes must support IPv4, IPv6, or both. A scan conducted on December 20, 2021, reveals the availability of 2527 probes with these attributes. Out of these, 1137 probes are IPv6 capable, while all of them support IPv4. These probes are distributed across 671 different Autonomous Systems (ASs) with varying densities in different regions: 70\% in Europe, 18\% in North America, 6\% in Asia, 3\% in Oceania, 1\% in Africa, and 1\% in South America. Before commencing the actual measurement series, which includes analyzing DNS resolvers from the edge and the core (as depicted in Figure \ref{fig:Intro}), examining DoTCP usage, EDNS(0) configuration, and DoTCP fallback, we evaluated 4343 probe resolvers associated with the 2443 participating probes, resulting in an average of 1.78 resolvers per probe.

% \usepackage{booktabs}
% \usepackage{graphicx}
\begin{table}[]
\centering
\caption{\em Probe distribution over Autonomous Systems along with the percentage of IPv4/IPv6 capable probes.}
\label{ASes_Probe_Distrbution}
\resizebox{\columnwidth}{!}{%
\begin{tabular}{@{}
>{\columncolor[HTML]{FFFFFF}}c 
>{\columncolor[HTML]{FFFFFF}}c 
>{\columncolor[HTML]{FFFFFF}}c 
>{\columncolor[HTML]{FFFFFF}}c 
>{\columncolor[HTML]{FFFFFF}}c @{}}
\toprule
\textbf{ASes} & \textbf{IPv4 capable} & \textbf{\% of probes} & \textbf{IPv6 capable} & \textbf{\% of probes} \\ \midrule
DTAG          & 137                   & 5.42\%                & 109                   & 9.59\%                \\
COMCAST       & 100                   & 3.96\%                & 61                    & 5.37\%                \\
VODANET       & 96                    & 3.80\%                & 48                    & 4.22\%                \\
PROXAD        & 89                    & 3.52\%                & 76                    & 6.68\%                \\
Orange S.A.   & 75                    & 2.97\%                & 58                    & 5.10\%                \\
AT\&T         & 51                    & 2.02\%                & 28                    & 2.46\%                \\
UUNET         & 37                    & 1.46\%                & 7                     & 0.62\%                \\
TNF-AS        & 34                    & 1.35\%                & 15                    & 1.32\%                \\
LDCOMNET      & 33                    & 1.31\%                & 10                    & 0.88\%                \\
KPN           & 30                    & 1.19\%                & 17                    & 1.50\%                \\ \bottomrule
\end{tabular}%
}
\end{table}

\subsection{From the edge}

For the purpose of analyzing the behavior of different DNS resolvers from the edge, we programmatically configured RIPE Atlas measurements specifically targeting the resolvers listed in Table \ref{Public_DNS_resolvers}. These measurements encompassed DNS resolution over both IPv4 and IPv6, utilizing both TCP and UDP transport protocols. In this study, we treated the DNS resolvers as black boxes, focusing on the direct communication between DNS client programs and recursive resolvers. Consequently, we did not examine the specific transport protocols or buffer sizes employed during communication with authoritative name servers. The measurements from the edge involved requesting "A" records for the widely-used domain "google.com". By choosing a heavily-cached domain, we aimed to minimize recursive resolution or unexpected errors, thereby emphasizing the communication between client programs and resolvers. The perceived response times by the probes should thus correspond to the duration of a simple UDP/TCP request and response. As mentioned earlier, the RIPE Atlas probes collected essential information during the measurements, including response times, error messages, and the UDP buffer sizes advertised by each resolver. This data was subsequently retrieved using the RIPE Atlas measurement API. To validate the collected response times and gain insights into the resolvers' distribution across the Internet, simultaneous Traceroute measurements were conducted. It is anticipated that the measured round-trip time (RTT) of a Traceroute to the resolver closely aligns with the response time of a DoUDP request when the queried domain is cached on the resolver.

\begin{figure*}[!t]
    \centering
    \includegraphics[height=9.81cm, width=1\linewidth]{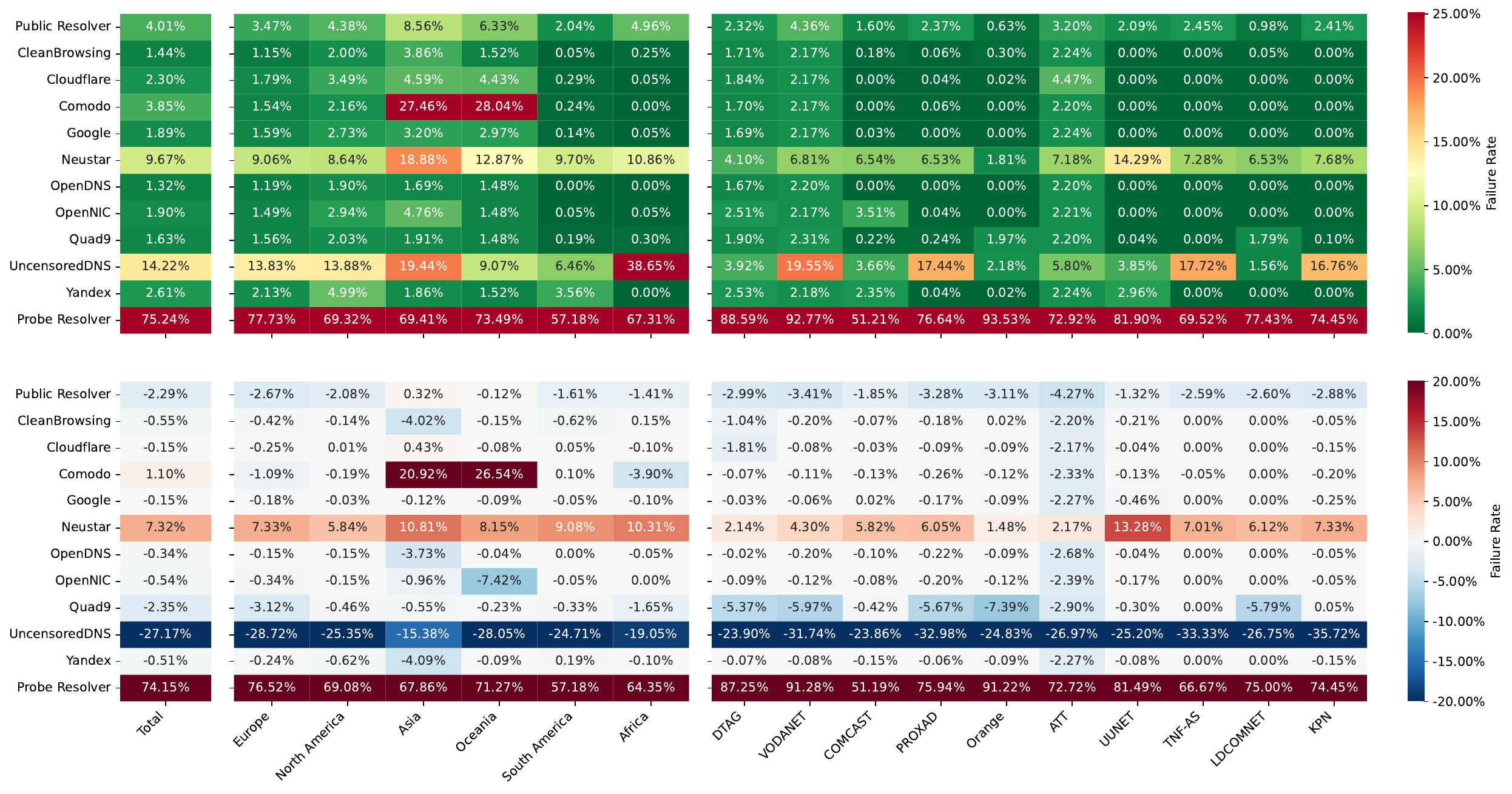}
    \caption{\em Failure rates observed from the edge over IPv4. The upper part represents the DoTCP failure rates of all resolvers in total and per continent and AS. The lower part reflects the difference between the DoTCP and the DoUDP failure rates for a particular pairing (a negative value hence indicates a higher DoUDP failure rate). 'Public Resolver' summarizes the observations of all resolvers that are not probe resolvers.}
    \label{fig:Image5}
\end{figure*}

\begin{figure*}[!t]
    \centering
    \includegraphics[height=9.81cm, width=1\linewidth]{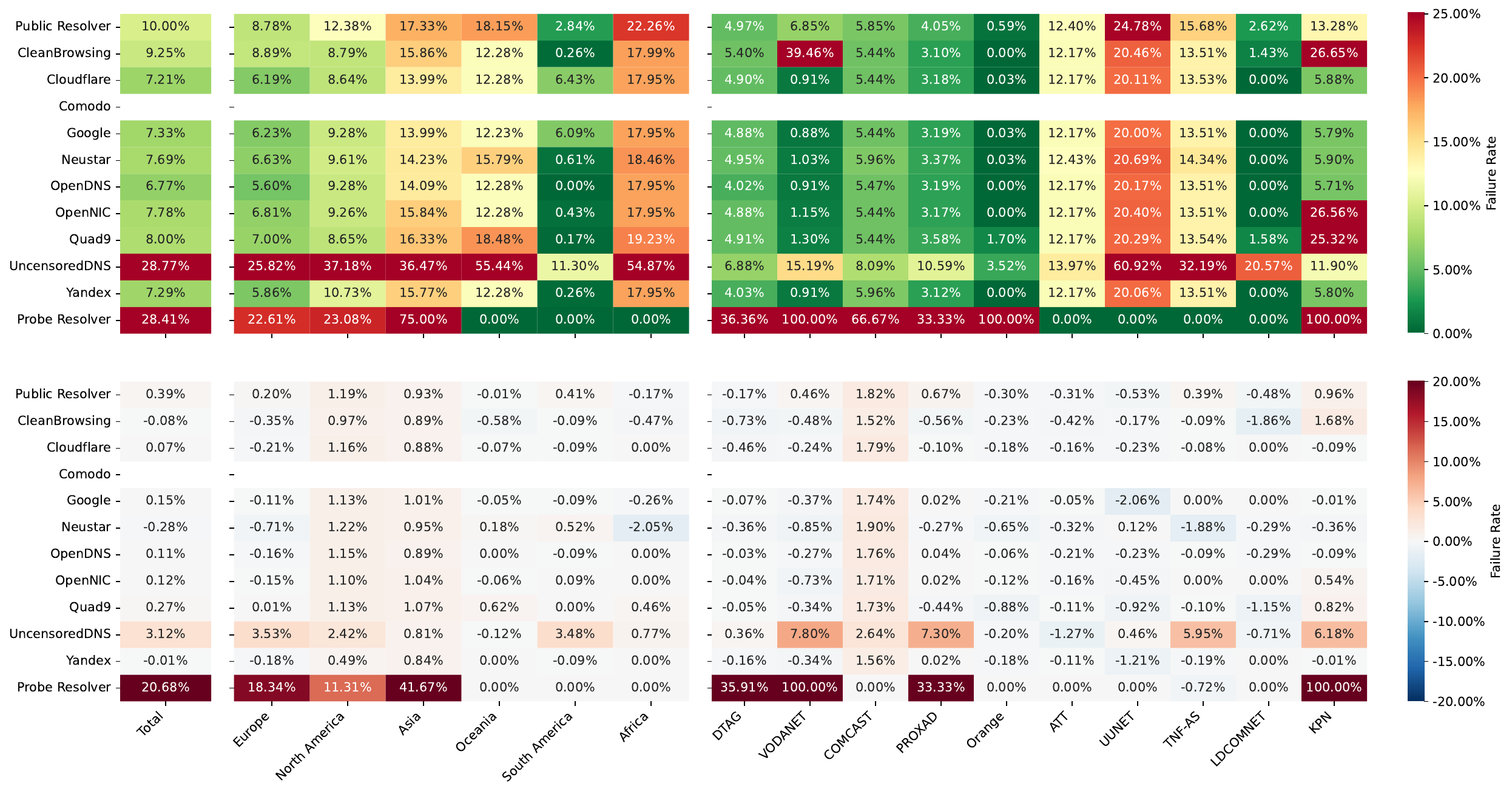}
    \caption{\em Failure rates observed from the edge over IPv6. The upper part presents the failure rates over DoTCP, the lower one the difference between DoTCP and DoUDP failure rates. White cells indicate that there is no data for the given pairing.}
    \label{fig:Image3}
\end{figure*}

\subsection{From the core}

This evaluation allows us to analyze resolver behavior when interacting with authoritative NS (see Figure \ref{fig:probe_distribution}). In this experiment, we use uncached domain names controlled by authoritative NS under our supervision. We analyze the resolvers' DNS configuration using two customized authoritative NSes. These NSes encode incoming DNS requests, including transport protocol and requester IP address, for later analysis. By observing the EDNS section of requests reaching the authoritative NS, we gain insights into the resolvers' EDNS configuration and potential usage of options like \textit{Cookie} or \textit{Client Subnet}.

\subsubsection{\textbf{DoTCP Usage and EDNS(0) Configuration}}
In the Core analysis, we utilize the RIPE Atlas network to evaluate the transport protocols and EDNS configuration employed by DNS resolvers. Our measurements focus on a target domain managed by our NSes to ensure uncached responses. To achieve uniqueness, each DNS request is modified with the probe ID and timestamp. By examining the IP addresses of the resolvers, we gain insights into their distribution across continents and AS networks. Additionally, observing the transport protocols used enables us to gather information on DoTCP usage in the Core.

\subsubsection{\textbf{DoTCP Fallback}}

The Core measurement focuses on observing the DoTCP fallback behavior of public DNS resolvers. Large responses, consisting of 72 AAAA records ($>$2KB response) for one server and 145 AAAA records ($>$4KB response) for the other, are returned by the authoritative NSes. By including different RR types (A, AAAA, and TXT) in each measurement, we aim to investigate the resolvers' reaction to both response sizes simultaneously. Given that the previous experiment revealed resolvers requesting both NSes equally, approximately 50\% of the requests are expected to receive 4KB and 2KB responses respectively. As large responses cannot be handled by UDP due to fragmentation issues, resolvers are anticipated to fallback to using DoTCP. The analysis focuses on whether the resolvers continue to utilize UDP or switch to DoTCP, providing insights into potential resiliency risks. Multiple requests from resolvers to the authoritative NS are expected, such as one over UDP followed by a fallback request over TCP. To accurately map incoming requests to the RIPE Atlas measurements, the domains queried by each probe are made unique using the aforementioned technique of prepending probe-specific information.

\section{Results}
\label{results}

We evaluate the results of the measurement from the edge concerning failure rates, response times, and EDNS(0) buffer sizes. Afterward, we analyze EDNS(0) configuration and  DoTCP fallback from the core.

\subsection{Probes}

A comprehensive examination of all RIPE Atlas probes reveals the presence of 2527 probes possessing the desired attributes. All probes exhibit compatibility with IPv4, while 1137 probes exhibit the additional capability of conducting measurements over IPv6. The geographic distribution of the probes, as well as the locations of the authoritative name servers, is visually depicted in Figure \ref{fig:probe_distribution}. Notably, a concentrated density of probes is observed in North America and Europe, which serves as the primary origin for the RIPE Atlas community. Specifically, Europe accounts for 70\% of the probes, followed by North America with 18\%, Asia with 6\%, Oceania with 3\%, Africa with 1\%, and South America with 1\%. The distribution of these probes among the Autonomous Systems is detailed in Table \ref{ASes_Probe_Distrbution}, highlighting the ten Autonomous Systems housing the majority of probes. The probes from Comcast, AT\&T, and UUNET are exclusively situated in North America, while the remaining Autonomous Systems are primarily distributed in Europe. Furthermore, it should be noted that certain Autonomous Systems mentioned in Table \ref{ASes_Probe_Distrbution} possess only a small number of IPv6-capable probes. The mapping between Autonomous System numbers and their respective names is obtained from IPtoASN\footnote{\label{1}\url{https://iptoasn.com}}.

\subsubsection*{\textbf{Probe Resolvers}}
\label{Probe resolvers}

Before conducting the actual measurement, preliminary test measurements are performed to gather address information regarding the locally configured resolvers on each probe. It is important to note that the resolvers can possess either publicly accessible IP addresses or private ones. For instance, a probe may have Google Public DNS configured, in addition to a DNS resolver exclusively accessible through its local network. Among the probes, 41.93\% have public IP addresses. Out of all the registered resolvers, 97.03\% are associated with their corresponding IPv4 addresses, while the remaining 2.97\% are linked to their IPv6 addresses. Additionally, it is crucial to acknowledge that during a RIPE Atlas measurement employing the utilized probe resolver parameter set, all probe resolvers are requested to resolve the relevant domain name. Consequently, both IPv4 and IPv6 DNS resolvers contribute to the measurement, regardless of the IP version mandated by the RIPE user.

\subsection{Evaluation from the edge}
\label{results from the edge}

\makeatletter
\newcolumntype{e}[1]{%--- Enumerated cells ---
   >{\minipage[t]{\linewidth}%
     \NoHyper%                Hyperref adds a vertical space
     \let\\\tabularnewline
    %\enumerate
      %  \addtolength{\rightskip}{0pt plus 50pt}% for raggedright
       % \setlength{\itemsep}{-\parsep}}%
       }
   p{#1}%
   <{\@finalstrut\@arstrutbox%\endenumerate
     \endNoHyper
     \endminipage}}

\newcolumntype{i}[1]{%--- Itemized cells ---
   >{\minipage[t]{\linewidth}%
        \let\\\tabularnewline
        %\itemize
           %\addtolength{\rightskip}{0pt plus 50pt}%
           %\setlength{\itemsep}{-\parsep}}%
           }
   p{#1}%
   <{\@finalstrut\@arstrutbox\endminipage}}

\begin{table}[!t]
\caption{\em EDNS(0) Buffer Sizes announced to the RIPE probes by the resolvers observed from the edge. %Buffer sizes that are not equal to 512, 1232, or 4096 bytes are summarized in the column \textit{other}. If EDNS is not used at all this is reflected in the column \textit{none}. 
CB: CleanBrowsing; U-DNS: UncensoredDNS}

\scalebox{0.73}
{
\begin{tabular}[t]{e{1.5cm}e{1.1cm}e{1.4cm}e{1.1cm}e{1.4cm}e{1.1cm}e{1.4cm}}

\hline
 &  &    \textbf{512} &   \textbf{1232} &   \textbf{4096} &   \textbf{none} &  \textbf{other} \\
 %\hline
\midrule
\rowcolor{gray!25} & \textit{IPv4} &  \cellcolor{magenta!40} \textbf{97.04\%} &   0.63\% &   1.46\% &   0.57\% &   0.30\% \\
\rowcolor{gray!25} \multirow{-2}{*}{\textbf{CB}} & \textit{IPv6} &  \cellcolor{magenta!40} \textbf{99.41\%} &   0.11\% &   0.48\% &   0.01\% &   0.00\% \\
 & \textit{IPv4}   &   0.20\% & \cellcolor{orange!40} \textbf{97.43\%} &   1.45\% &   0.53\% &   0.40\% \\
\multirow{-2}{*}{\textbf{Cloudflare}}  & \textit{IPv6} &   0.11\% &  \cellcolor{orange!40} \textbf{99.44\%} &   0.44\% &   0.01\% &   0.00\% \\
\rowcolor{gray!25} & \textit{IPv4}   &   0.18\% &   0.64\% &  \cellcolor{cyan!40} \textbf{98.30\%} &   0.57\% &   0.30\% \\
\rowcolor{gray!25} \multirow{-2}{*}{\textbf{Comodo}}  & \textit{IPv6} &  - &   - &  \cellcolor{cyan!40} - &   - &  - \\
 & \textit{IPv4}  &  \cellcolor{magenta!40} \textbf{96.82\%} &   0.78\% &   1.47\% &   0.58\% &   0.34\% \\
 \multirow{-2}{*}{\textbf{Google}}  & \textit{IPv6} &  \cellcolor{magenta!40} \textbf{99.22\%} &   0.10\% &   0.67\% &   0.00\% &   0.01\% \\
\rowcolor{gray!25} & \textit{IPv4}  &   0.18\% &   0.64\% &  \cellcolor{cyan!40} \textbf{98.32\%} &   0.56\% &   0.30\%\\
\rowcolor{gray!25} \multirow{-2}{*}{\textbf{Neustar}} & \textit{IPv6} &   0.10\% &   0.10\% & \cellcolor{cyan!40} \textbf{99.79\%} &   0.00\% &   0.00\% \\
 & \textit{IPv4} &   0.18\% &   0.63\% &  \cellcolor{cyan!40} \textbf{98.20\%} &   0.57\% &   0.43\% \\
 \multirow{-2}{*}{\textbf{OpenDNS}}      & \textit{IPv6} &   0.10\% &   0.11\% & \cellcolor{cyan!40} \textbf{99.79\%} &   0.00\% &   0.00\% \\
\rowcolor{gray!25}     & \textit{IPv4}  &   0.18\% & \cellcolor{orange!40} \textbf{97.53\%} &   1.42\% &   0.56\% &   0.30\% \\
\rowcolor{gray!25} \multirow{-2}{*}{\textbf{OpenNIC}} & \textit{IPv6} &   0.11\% &  \cellcolor{orange!40} \textbf{99.43\%} &   0.47\% &   0.00\% &   0.00\% \\
  & \textit{IPv4}  &  \cellcolor{yellow!40} \textbf{19.15\%} &  \cellcolor{yellow!40} \textbf{55.47\%} &   1.48\% &  \cellcolor{yellow!40} \textbf{23.55\%} &   0.35\% \\
 \multirow{-2}{*}{\textbf{Quad9}} & \textit{IPv6} &  \cellcolor{yellow!40} \textbf{20.98\%} &  \cellcolor{yellow!40} \textbf{62.09\%} &   0.47\% &  \cellcolor{yellow!40} \textbf{16.46\%} &   0.00\% \\
\rowcolor{gray!25} & \textit{IPv4} &   0.30\% &  \cellcolor{orange!40} \textbf{95.87\%} &   2.39\% &   0.96\% &   0.49\% \\
\rowcolor{gray!25} \multirow{-2}{*}{\textbf{U-DNS}} & \textit{IPv6} &   0.13\% &  \cellcolor{orange!40} \textbf{99.29\%} &   0.57\% &   0.01\% &   0.00\% \\
& \textit{IPv4} &   0.19\% &   0.64\% &  \cellcolor{cyan!40} \textbf{98.04\%} &   0.75\% &   0.39\% \\
 \multirow{-2}{*}{\textbf{Yandex}}  & \textit{IPv6} &   0.11\% &   0.10\% &  \cellcolor{cyan!40} \textbf{99.79\%} &   0.00\% &   0.00\% \\ \hline
\rowcolor{gray!25}           & \textit{IPv4} &  24.97\% &  36.12\% &  35.30\% &   3.24\% &   0.36\% \\
\rowcolor{gray!25} \multirow{-2}{*}{\textbf{Overall}} & \textit{IPv6}           &  24.86\% &  38.81\% &  34.46\% &   1.87\% &   0.00\% \\
%\hline
\bottomrule
\end{tabular}
%\end{center}
}

\label{table:EDGEfrontendbuffersizes}
\end{table}

\begin{figure*}[!t]
    \centering
    \includegraphics[height=10.1cm, width=1\linewidth]{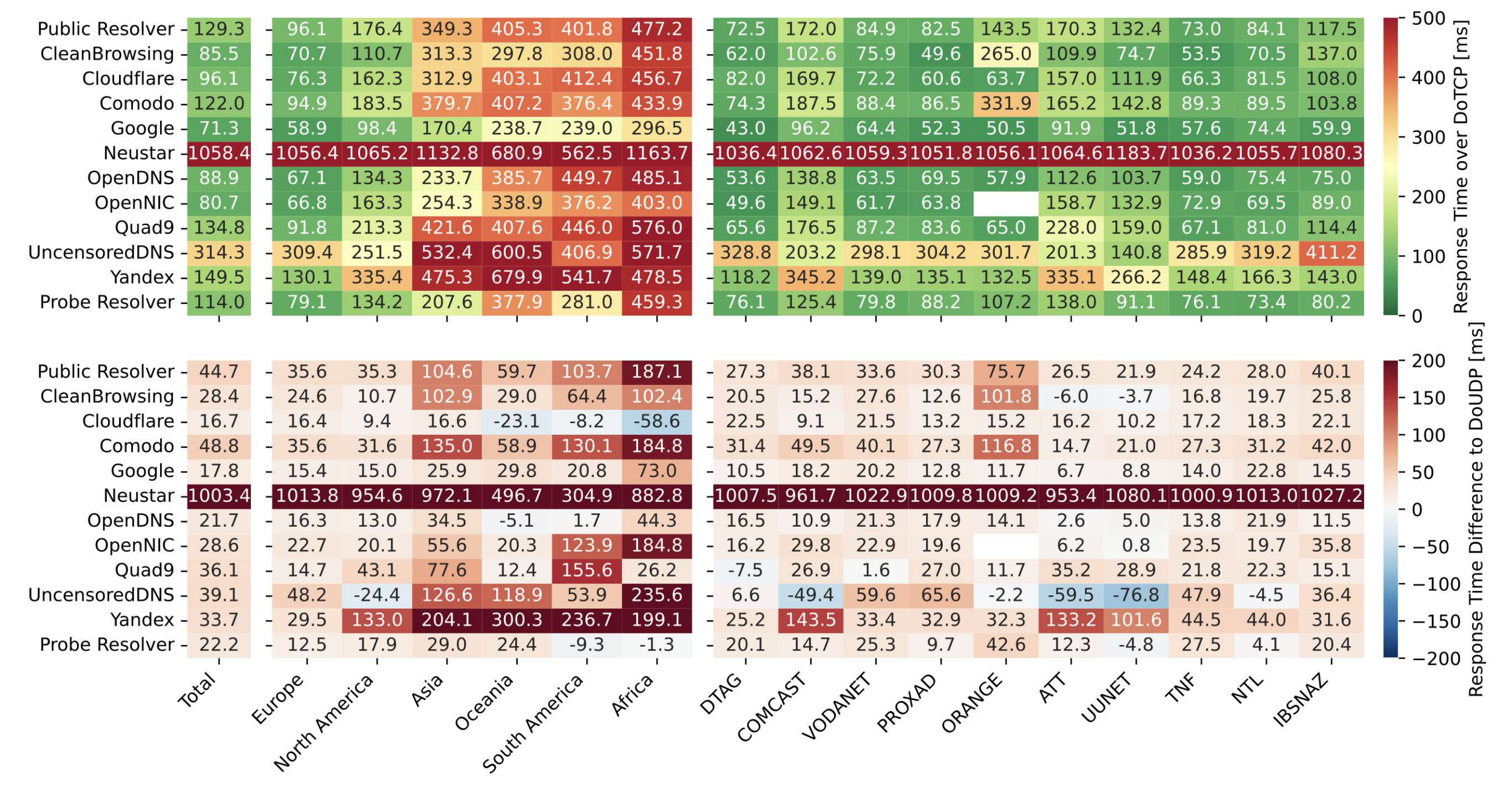}
    \caption{\em Response times (RTs) observed from the edge over IPv4. The values represent the median RT over the medians of each probe. The upper part shows the results for DoTCP, the lower one the differences between DoTCP and DoUDP.}
    \label{fig:Response_Times_IPv4}
   
\end{figure*}

\begin{figure*}[!t]
    \centering
    \includegraphics[height=10.1cm, width=1\linewidth]{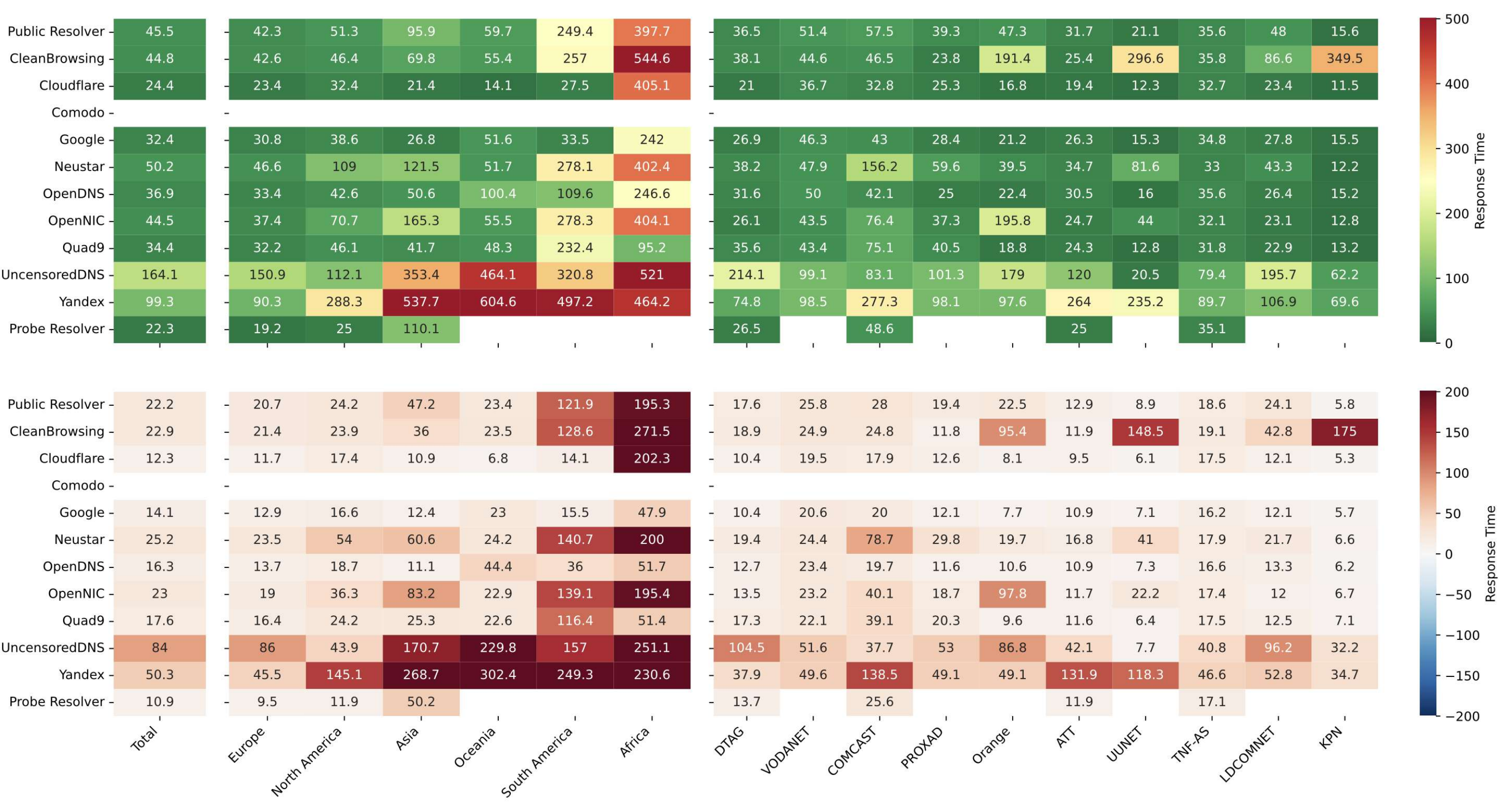}
    \caption{\em Response times (RTs) observed from the edge over IPv6. The values represent the median RT over the medians of each probe. The upper part shows the results for DoTCP, and the lower one the differences between DoTCP and DoUDP. White cells indicate that there is no data for the given pairing.}
    \label{fig:Response_Times_IPv6}
    
\end{figure*}

\begin{figure*}[!t]
     \centering
\begin{subfigure}{0.45\linewidth}
    \includegraphics[height=3cm, width=1\linewidth]{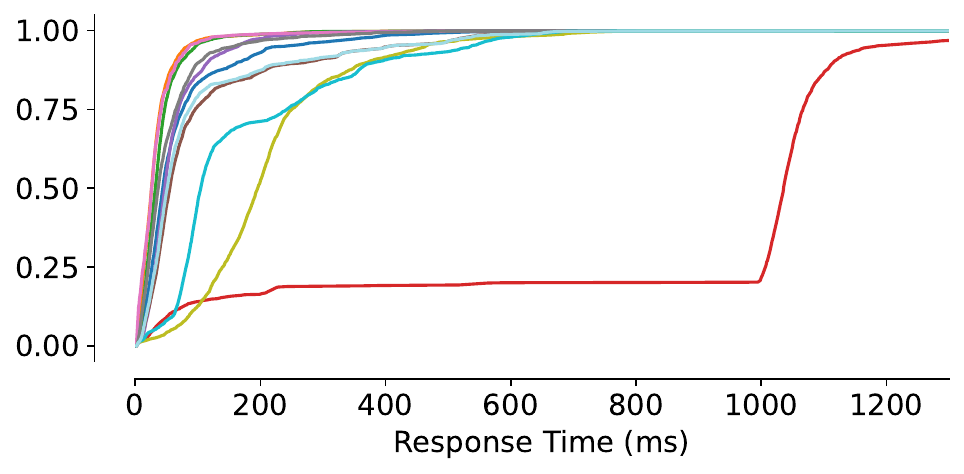}
    \caption{Response times over DoTCP and IPv4}
    \label{DoTCP and IPv4}
\end{subfigure}
\hfill
\begin{subfigure}{0.45\linewidth}
    \includegraphics[height=3cm, width=1\linewidth]{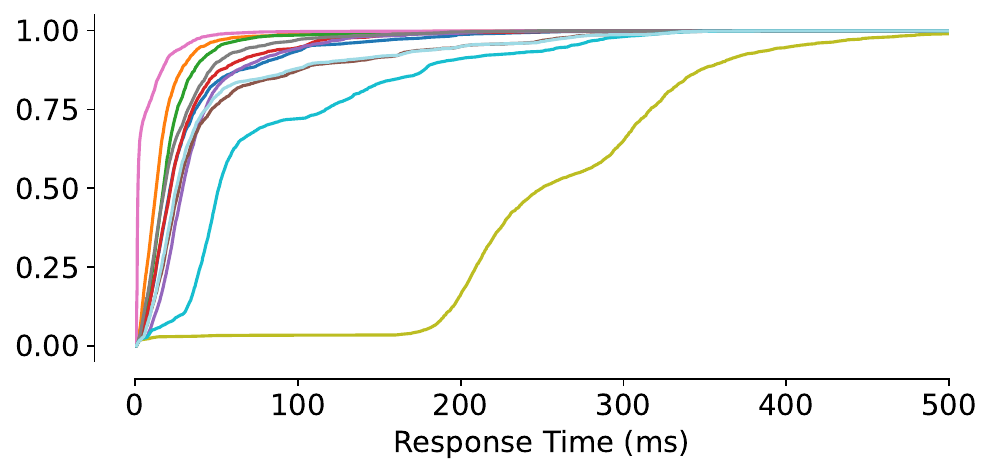}
    \caption{Response times over DoUDP and IPv4}
    \label{DoUDP and IPv4}
\end{subfigure}
\hfill
\begin{subfigure}{0.45\linewidth}
    \includegraphics[height=3cm, width=1\linewidth]{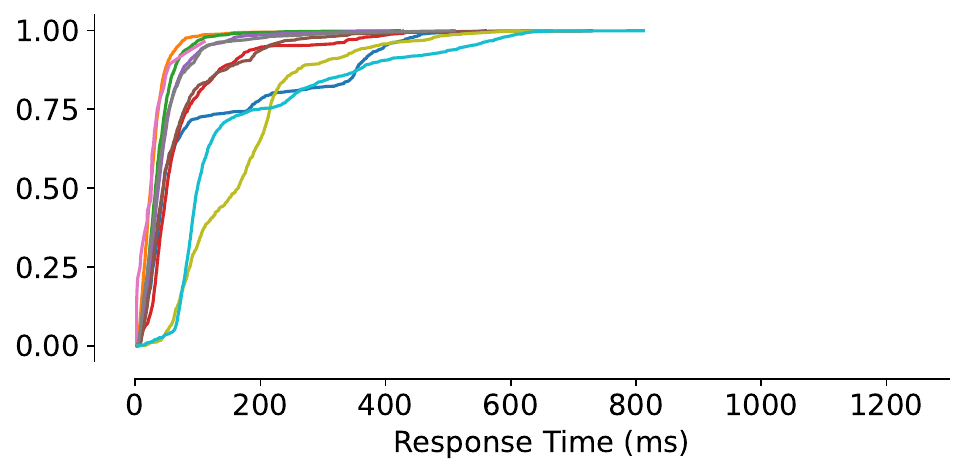}
    \caption{Response times over DoTCP and IPv6}
    \label{DoTCP and IPv6}
\end{subfigure}
\hfill
\begin{subfigure}{0.45\linewidth}
    \includegraphics[height=3cm, width=1\linewidth]{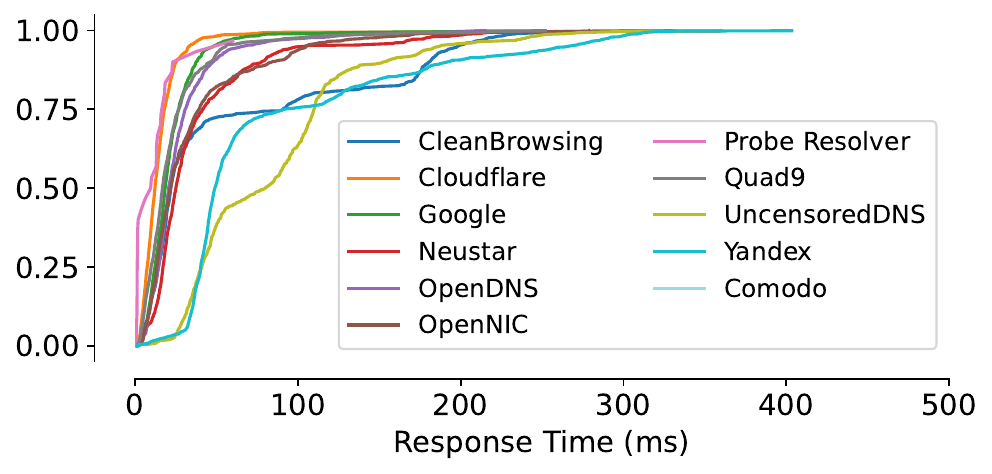}
    \caption{Response times over DoUDP and IPv6}
    \label{DoUDP and IPv6}
\end{subfigure}
\caption{\em Response times of each resolver over TCP/UDP and IPv4/IPv6 as CDF. The curves present an accumulation of the medians RTs of each probe.}

\label{combinationTCP_UDP_IPv4_IPv6}

\end{figure*}

%To conduct edge analysis, we performed a week-long data collection period. Each day, we carried out ten identical measurement intervals, each spanning two hours. Our measurements focused on the domain "google.com" and utilized RIPE Atlas DNS measurements targeting both Public and Probe resolvers. We employed the IPv4 and IPv6 protocols, as well as UDP and TCP transport protocols. All available probes capable of the respective IP version were enlisted for participation. Overall, we obtained 4,969,141 individual results from the participating probes. Specifically, there were 1,826,347 results for DoTCP over IPv4, 1,826,923 results for DoUDP over IPv4, 658,743 results for DoTCP over IPv6, and 657,128 results for DoUDP over IPv6. In addition to the DNS measurements, we conducted daily measurement intervals for Traceroute. These measurements targeted both Public and Probe resolvers with publicly accessible IP addresses, utilizing both IPv4 and IPv6. We successfully collected a total of 163,867 individual Traceroutes over IPv4 and 28,643 over IPv6.

This section analyzes failure rates of transport protocols at the edge to assess DNS resilience. Response times of resolvers are compared and the performance of public resolvers is evaluated using DNS response times and traceroute round-trip times. The adoption of DNS Flag Day recommendations by individual resolvers is analyzed through EDNS(0) buffer size announced to RIPE probes.

\subsubsection{\textbf{Failure Rates}}
\label{edge_failure rate}

Based on Kosek \emph{et al.}'s research in \cite{DoTCP_Kosek}, failed measurements are defined as those with no DNS response at the probe. In IPv4, public resolvers show lower failure rates for DoTCP (4.01\%) compared to DoUDP (6.3\%), indicating higher resiliency of DoTCP (Figure \ref{fig:Image5}). However, probe resolvers present a different scenario with DoTCP surpassing DoUDP by 74.15\%. DoUDP failures are solely due to Timeouts (5000ms). And public resolvers' DoTCP failures are also primarily caused by Timeouts (42.75\%), READ-ERROR (33.91\%), CONNECT-ERROR (23.24\%), and TCP-READ (0.09\%). Bad address (99.17\%) is the main cause of DoTCP failures for probe resolvers. Overall, probe resolvers exhibit significantly higher DoTCP failure rates across continents. In case of IPv6, for public resolvers, we find lower resiliency over both transport protocols (DoTCP 10\%, DoUDP 9.61\%). Most public resolvers exhibit failure rates between 6.77\% and 9.25\%, see Figure \ref{fig:Image3}. Uncensored DNS shows by far the worst DoTCP and DoUDP resiliency.

To analyze the adoption of the DNS Flag Day 2020 recommendations by the public resolvers from the edge, we evaluate the EDNS(0) buffer sizes which the individual resolvers announce to the RIPE Atlas probes. Table \ref{table:EDGEfrontendbuffersizes} summarizes the buffer sizes that have been observed in the UDP measurements. As for all resolvers except Quad9 the difference in the percentages of the announced buffer sizes between IPv4 and IPv6 are fairly low ($\le$3.5\%). The buffer sizes advertised by Cloudflare, OpenNIC, UncensoredDNS, and Quad9 (55.47\% IPv4, 62.09\% IPv6) conform to the DNS Flag Day 2020 recommendation of a default buffer size of 1232B in most cases. Neustar, Comodo, OpenDNS and Yandex mainly use 4096 bytes. In 23.55\% of the Quad9 DNS responses over IPv4, EDNS(0) is not used at all leaving clients to the default DoUDP message size limit of 512 bytes. This first view from the edge shows that several public DNS resolvers still lack adoption to the DNS Flag Day 2020 recommendations. To see whether this also holds for the communication with authoritative NSes, we conducted another experiment from the core.

\subsubsection{\textbf{Response Times}}
\label{edge_response times results}
RIPE Atlas employs a measurement methodology to assess the response time (RT) of DNS requests. It measures the duration from the initiation of the measurement until a valid DNS response is received at the probe. DoUDP enables immediate transmission of requests to resolvers without the need to establish a TCP connection like DoTCP. Considering the cached nature of the "google.com" domain, DoUDP requests to resolvers with efficient cache management are expected to have response times equivalent to the probe-resolver round-trip time. In contrast, DoTCP measurements involve a three-way handshake, resulting in response times roughly twice as long as DoUDP. To ensure a fair comparison, only probe-resolver pairs with successful responses over both TCP and UDP are considered. The depicted response times in Figures \ref{fig:Response_Times_IPv4} and \ref{fig:Response_Times_IPv6} are obtained through a two-tiered approach, calculating the median response time for each probe-resolver combination and presenting the median of all probes.

\textbf{IPv4.}
The analysis of response times for Public and Probe resolvers confirms the expected pattern of approximately doubled median response times for DoTCP compared to DoUDP. Probe resolvers, due to their close physical proximity to the probing device, exhibit faster response times than Public resolvers. %Notably, 58.07\% of the Probe resolvers have private IP addresses, indicating their extremely close proximity. 
Among the Public DNS resolvers examined, Cloudflare, Google, and Quad9 demonstrate the lowest median response times for both transport protocols. Yandex shows relatively long response times for both DoTCP (104.5ms) and DoUDP (51.3ms). Neustar, on the other hand, exhibits the longest median DoTCP response time (1035.8ms) in this experiment, along with high DoTCP failure rates, suggesting inadequate implementation. %UncensoredDNS displays comparatively long response times over both transport protocols, and notably, its UDP response times are slower than TCP response times. 
When comparing response times by continent, Public resolvers generally respond fastest to DoTCP and DoUDP requests from European (DoTCP 52.7ms, DoUDP 24.1ms) and North American (DoTCP 54.9ms, DoUDP 27.1ms) probes. %UncensoredDNS shows high variability in DoUDP response times across continents. Neustar consistently displays extremely high DoTCP response times (over 1 second) from all continents. 

\begin{figure*}[!t]
\centering
\begin{subfigure}{0.45\linewidth}
    \includegraphics[height=3cm, width=1\linewidth]{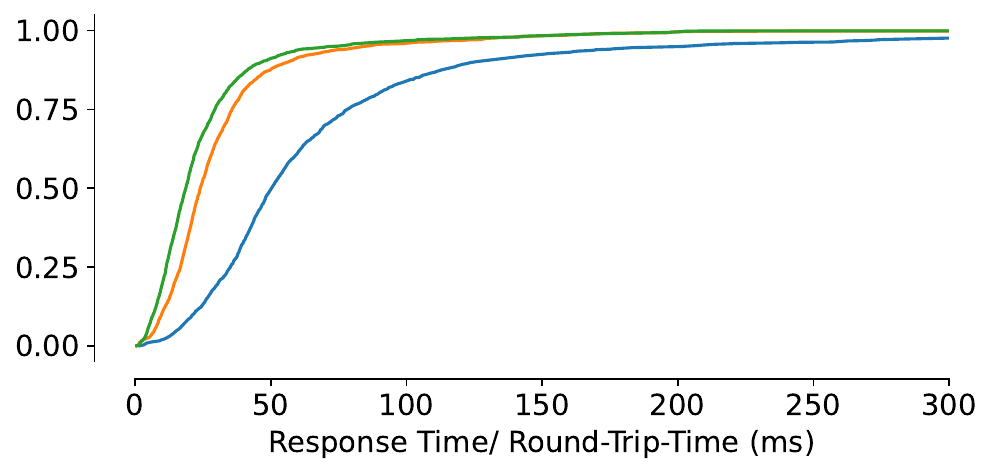}
    \caption{IPv4}
    \label{IPv4}
\end{subfigure}
\hfill
\begin{subfigure}{0.45\linewidth}
    \includegraphics[height=3cm, width=1\linewidth]{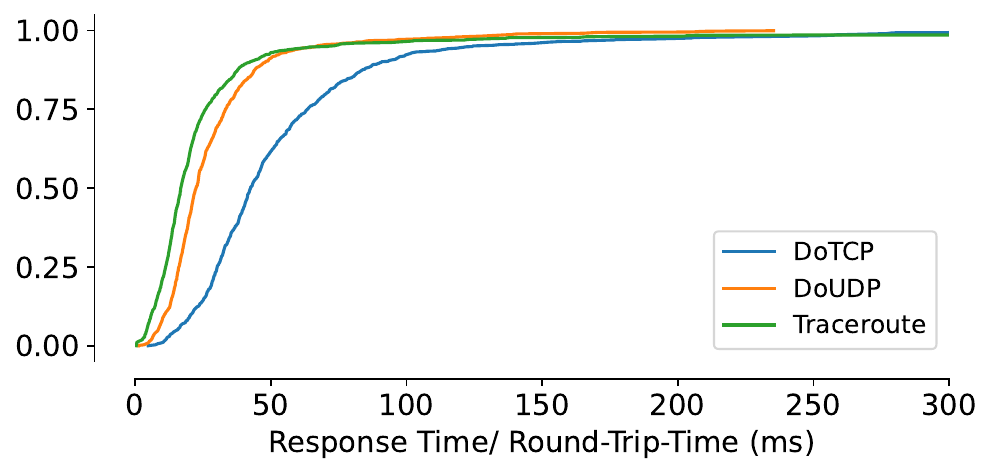}
    \caption{IPv6}
    \label{IPv6}
\end{subfigure}

\caption{\em Traceroute round-trip-times (RTTs) and DNS response times (RTs) as a CDF. Again, the curves reflect an accumulation of probe-medians.}
\label{Traceroute_RTT_IPv4_IPv6}

\end{figure*}

The increased response times observed across various continent/resolver combinations for both transport protocols can be attributed to the sparser distribution of resolver Points-of-Presence (PoPs) in those continents. This is evident as DoTCP response times are consistently approximately twice as high as DoUDP response times, indicating longer Round-Trip Time (RTT) due to greater distances between probes and resolvers. Probe resolvers demonstrate relatively low response times for both DoTCP (6.4ms - 32.2ms) and DoUDP (3.2ms - 15.3ms). Among Public resolvers, Orange Autonomous System (AS) exhibits the highest overall median DoTCP response time (167.2ms) and also shows elevated DoTCP response times for CleanBrowsing, Comodo, and OpenNIC resolvers. Overall, most ASes, primarily operating in North America and Europe, do not exhibit significant anomalies in DoTCP and DoUDP response times. %However, Yandex and UncensoredDNS show substantial variations in DoTCP performance across different ASes. 
Yandex performs poorly for requests from North American ASes (223.9-290ms) but slightly better for European ASes (71.9-103.6ms). Conversely, UncensoredDNS displays higher DoTCP response times for European ASes (117.8-214.6ms) compared to North American ASes (36.6-145.9ms). Notably, UncensoredDNS consistently exhibits higher DoUDP response times than DoTCP, including a median DoUDP response time over seven times higher for requests from UUNET. Furthermore, %Neustar consistently displays response times exceeding 1 second for all examined ASes.

\textbf{IPv6.} The majority of Public resolvers show similar median response times for both DoTCP and DoUDP compared to their IPv4 counterparts. However, there is a notable decrease in the median response time for DoTCP requests to Neustar (50.2ms) over IPv6. UncensoredDNS and Yandex still exhibit relatively high DoTCP response times. The improved performance of Neustar over DoTCP and UncensoredDNS over DoUDP contributes to lower overall median response times for public resolvers across both transport protocols. It is important to mention that probe resolvers are not considered in the IPv6 analysis. The analysis based on continents for Public resolvers reveals results similar to those observed in IPv4 measurements. Notably, the DoTCP response time for Africa is more than 70ms higher in IPv6 compared to IPv4. This can be attributed to the relatively poorer DoTCP performance of Cloudflare, Google, and OpenDNS resolvers for requests originating from Africa over IPv6. Analyzing DNS response times over IPv6 by Autonomous System reveals higher variations compared to IPv4. Generally, probes from UUNET (21.1ms) and KPN (15.6ms) receive the fastest DoTCP responses, with most Public resolvers displaying their minimal DoTCP response times for these two ASes. CleanBrowsing shows outliers in DoTCP response times for probes from the Orange (191.4ms), UUNET (296.6ms), and KPN (349.5ms). UncensoredDNS shows relatively high response times for requests from all Autonomous Systems (62.2ms-214.1ms) except for UUNET (20.5ms).

\begin{summary}
\takeaway{In IPv4, Cloudflare and Google DNS resolvers demonstrate the most stable DoTCP and DoUDP response times across all continents. Other Public resolvers generally have significantly higher DoTCP response times for at least one, and often multiple, continents, particularly in Africa (322.5ms) and South America (245.8ms) where they take the longest to respond. Similar patterns are observed for DoUDP response times. Whereas, in IPv6 measurements, Europe exhibits the lowest median response times for DoTCP and DoUDP, while Africa exhibits the highest. Cloudflare seems to have fewer IPv6-capable Points-of-Presence distributed in Africa, resulting in increased round-trip times (RTTs) due to long distances.}

\end{summary}

\begin{figure*}[!t]
     \centering
\begin{subfigure}{0.45\textwidth}
    \includegraphics[height=3cm, width=1\linewidth]{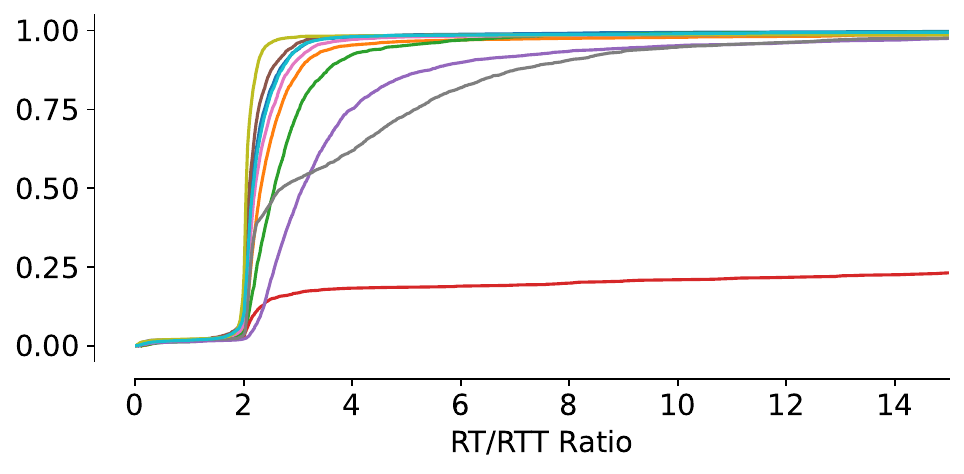}
    \caption{DoTCP IPv4}
    \label{DoTCP IPv4}
\end{subfigure}
\hfill
\begin{subfigure}{0.45\textwidth}
    \includegraphics[height=3cm, width=1\linewidth]{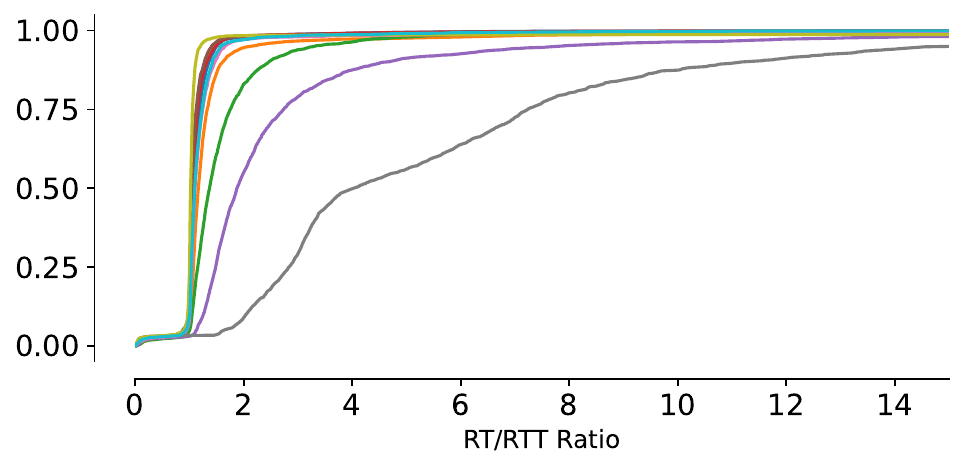}
    \caption{DoUDP IPv4}
    \label{DoUDP IPv4}
\end{subfigure}

\begin{subfigure}{0.45\textwidth}
    \includegraphics[height=3cm, width=1\linewidth]{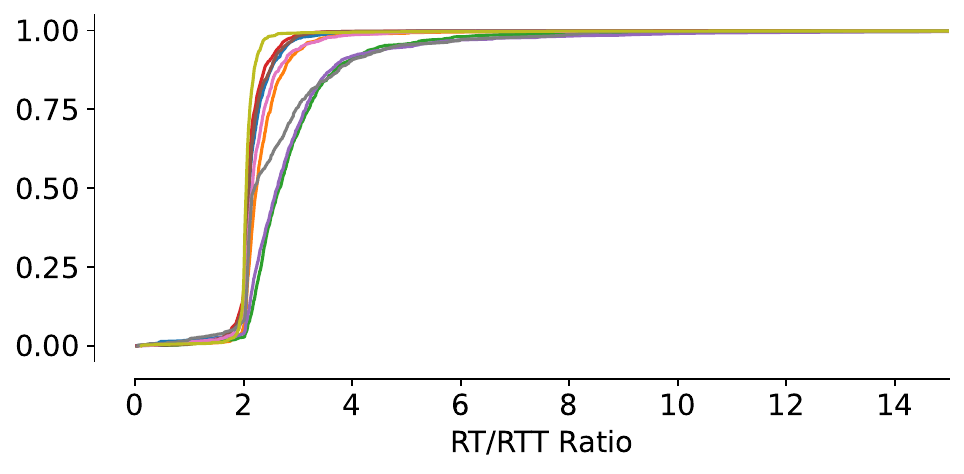}
    \caption{DoTCP IPv6}
    \label{DoTCP IPv6}
\end{subfigure}
\hfill
\begin{subfigure}{0.45\textwidth}
    \includegraphics[height=3cm, width=1\linewidth]{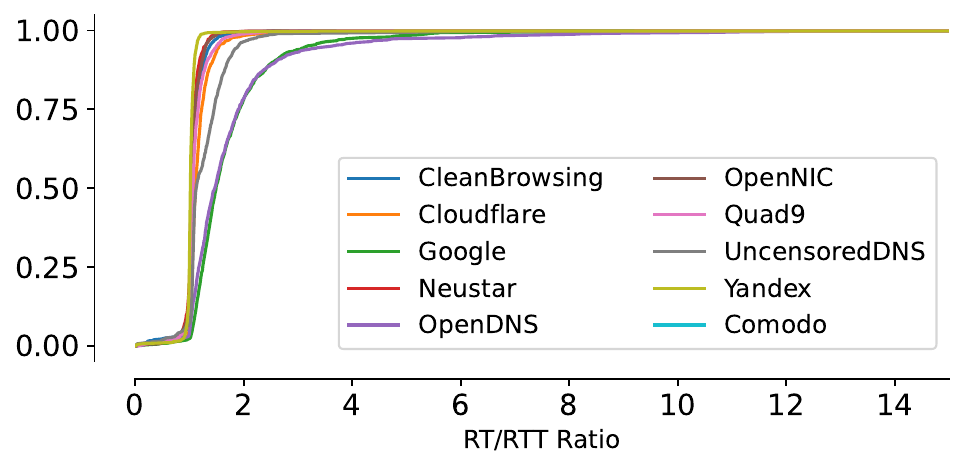}
    \caption{DoUDP IPv6}
    \label{DoUDP IPv6}
\end{subfigure}

\caption{ \em RT-RTT ratio (quotient of median round-trip-time and median response time for each probe) for each resolver over TCP/UDP and IPv4/IPv6 as CDF.}
\label{RT_RTT_Ratio}

\end{figure*}

\textbf{Combination - IPv4 and IPv6.} The evaluation of the measured response times of DNS over both transport protocols and IP versions as CDF emphasizes the results summarized above and allows their comparison from a different standpoint. Instead of the median aggregated median RTs of each probe shown in Figure [\ref{fig:Response_Times_IPv4}-\ref{fig:Response_Times_IPv6}], Figure \ref{combinationTCP_UDP_IPv4_IPv6} shows the accumulation of all recorded probe-medians based on the transport protocol and IP version used. the Figure \ref{DoTCP and IPv4} confirms the extremely high DoTCP response times of Neustar DNS over IPv4. It shows that around 80\% of the DoTCP requests take more than one second to be answered. The comparably bad performance of Yandex and UncensoredDNS can be seen for all combinations of transport protocol and IP version, especially
for UncensoredDNS over UDP and IPv4 (more than 90\% of the requests have an RT of more than 150ms). Furthermore, the response times of CleanBrowsing over both transport protocols on average increase when IPv6 is used instead of IPv4. This observation is not reflected by the median response times as the effect shows mainly for the slowest 25\% of requests. Figure \ref{combinationTCP_UDP_IPv4_IPv6} also confirms that Cloudflare and Google exhibit the most stable DNS response times of all Public resolvers.

\subsubsection{\textbf{Traceroute}}
\label{Traceroute}

RIPE Atlas does not allow Traceroute measurements within the probe networks, and the focus of this paper is primarily on the performance, resilience, and security of Public resolvers while Probe resolvers are excluded from the subsequent analysis. Ideally, the resolvers should exhibit DoUDP response times that are roughly similar to the RTTs measured through Traceroute; however, not exactly the same, as ICMP (used in Traceroute) requires slightly less overhead for transportation than UDP. Figure \ref{Traceroute_RTT_IPv4_IPv6} presents the cumulative distribution function (CDF) of median response times and round-trip times for all Public resolvers over each probe, validating these expectations for both IPv4 and IPv6. %While DoUDP response times and observed RTTs are highly similar, DoTCP response times are higher. This suggests that many resolvers require additional overhead for domain name resolution over TCP. 
DoTCP over IPv6 aligns more closely with the anticipated ratio of response times to round-trip times. The response time/round-trip time ratio (RT/RTT ratio) represents the quotient of the median response time and the median RTT between the probe and the resolver. Ideally, resolvers should have an RT/RTT ratio of 1 for DoUDP and 2 for DoTCP. Figure \ref{RT_RTT_Ratio} presents the RT/RTT ratios per resolver, transport protocol, and IP version. For Neustar over DoTCP and IPv4, the observed ratio exceeds 2 for over 80\% of the probes. This further highlights that the very high response times of the resolver are not due to a sparsely distributed global network of Points-of-Presence but rather an inadequate DoTCP implementation at various PoPs. The same conclusion can be drawn for UncensoredDNS over both TCP and UDP, as shown in Figures \ref{RT_RTT_Ratio}. OpenDNS and Google exhibit higher RT/RTT ratios compared to other Public resolvers over both transport protocols and IP versions, indicating that their relatively fast DNS response times are more attributed to a well-distributed global network rather than exceptionally efficient request processing.

\begin{figure*}[!t]
    \centering
    \includegraphics[height=9.81cm,width=1\linewidth]{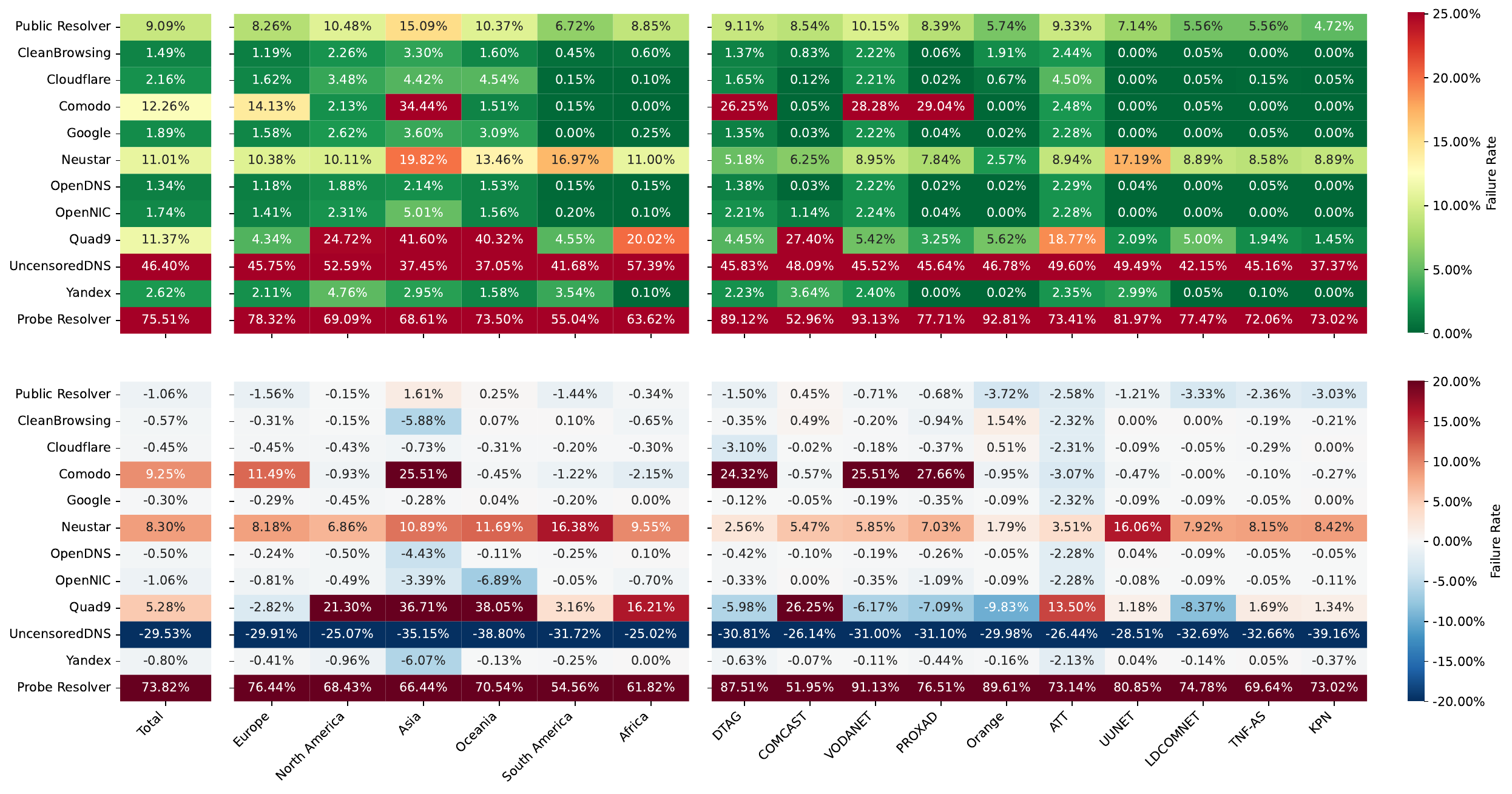}
    \caption{\em Failure rates observed from the core over IPv4. The upper part represents the DoTCP failure rates of all resolvers in total and per continent and AS. The lower part reflects the difference between the DoTCP and the DoUDP failure rates for a particular pairing (a negative value hints at a higher DoUDP failure rate). Public Resolver summarizes the observations of all resolvers that are not Probe resolvers.}
    \label{fig:Image7}
\end{figure*}

\begin{figure*}[!t]
    \centering
    
    \includegraphics[height=9.81cm,width=1\linewidth]{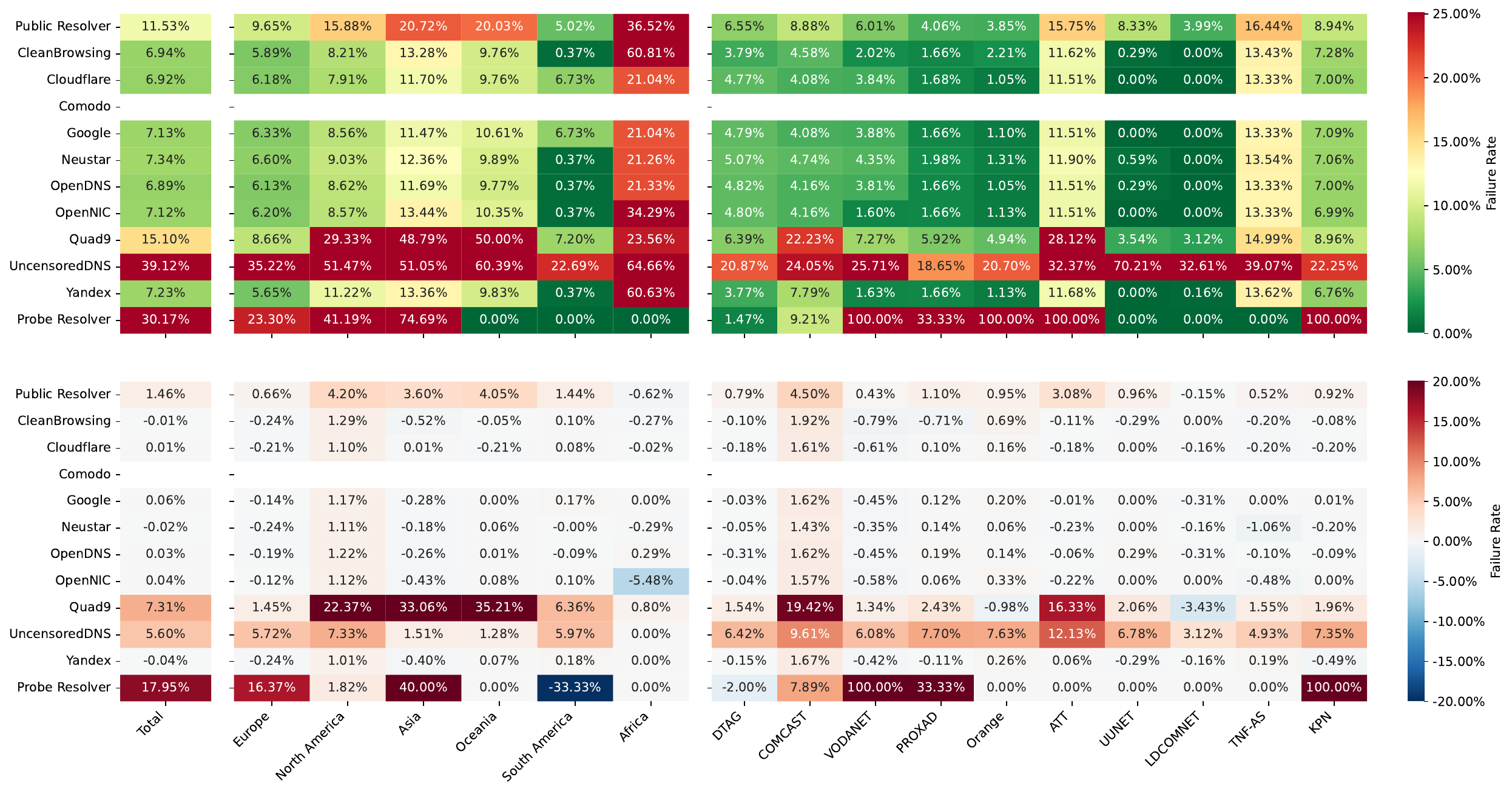}
    \caption{\em Failure rates observed from the core over IPv6. The upper part presents the failure rates over DoTCP, and the lower one is the difference between DoTCP and DoUDP failure rates. White cells indicate that there is no data for the given pairing.}
    \label{fig:Image8}
\end{figure*}

\begin{table}
\caption{\em EDNS(0) Buffer Sizes announced to RIPE probes by public resolvers in the measurement series from the core. Buffer sizes that are not equal to 512, 1232, or 4096 bytes are summarized in the column \textit{other}. If EDNS is not used at all this is reflected in the column \textit{none}.}
\begin{center}
\resizebox{\columnwidth}{!}{%
\begin{tabular}{lrrrrrr}
\toprule
 &  &    \textbf{512} &   \textbf{1232} &   \textbf{4096} &   \textbf{none} &  \textbf{other} \\
\midrule
\rowcolor{gray!25} & \textit{IPv4} & \cellcolor{magenta!40} \textbf{97.83\%} &   1.04\% &   0.56\% &  0.31\% &   0.27\% \\
\rowcolor{gray!25} \multirow{-2}{*}{\textbf{CleanBrowsing}} & \textit{IPv6} & \cellcolor{magenta!40} \textbf{99.33\%} &   0.19\% &   0.20\% &  0.29\% & 0.00\%\\   
 & \textit{IPv4}  &   0.47\% &  \cellcolor{orange!40} \textbf{98.24\%} &   0.61\% &  0.31\% &   0.38\% \\
\multirow{-2}{*}{\textbf{Cloudflare}}  & \textit{IPv6}    &   0.22\% &  \cellcolor{orange!40} \textbf{99.29\%} &   0.20\% &  0.28\% & 0.00\% \\
\rowcolor{gray!25} & \textit{IPv4} &   0.45\% &   1.02\% &  \cellcolor{cyan!40} \textbf{97.94\%} &  0.31\% &   0.27\% \\
\rowcolor{gray!25} \multirow{-2}{*}{\textbf{Comodo}}  & \textit{IPv6}  &  - &   - &   \cellcolor{cyan!40} - &  - & - \\
 & \textit{IPv4} & \cellcolor{magenta!40} \textbf{97.61\%} &   1.20\% &   0.53\% &  0.30\% &   0.35\% \\
 \multirow{-2}{*}{\textbf{Google}}  & \textit{IPv6}  &  \cellcolor{magenta!40} \textbf{99.23\%} &   0.18\% &   0.31\% &  0.29\% & 0.00\% \\
\rowcolor{gray!25} & \textit{IPv4}  &   0.45\% &   0.98\% &  \cellcolor{cyan!40} \textbf{97.99\%} &  0.31\% &   0.27\% \\
\rowcolor{gray!25} \multirow{-2}{*}{\textbf{Neustar}} & \textit{IPv6}  &   0.24\% &   0.18\% &  \cellcolor{cyan!40} \textbf{99.29\%} &  0.29\% & 0.00\% \\
& \textit{IPv4}  &   0.54\% &   0.98\% &  \cellcolor{cyan!40} \textbf{97.86\%} &  0.31\% &   0.31\% \\
 \multirow{-2}{*}{\textbf{OpenDNS}}      & \textit{IPv6}  &   0.24\% &   0.18\% &  \cellcolor{cyan!40} \textbf{99.30\%} &  0.29\% & 0.00\% \\
\rowcolor{gray!25}     & \textit{IPv4}  &   0.46\% &  \cellcolor{orange!40} \textbf{98.41\%} &   0.56\% &  0.31\% &   0.27\% \\
\rowcolor{gray!25} \multirow{-2}{*}{\textbf{OpenNIC}} & \textit{IPv6} &   0.24\% &  \cellcolor{orange!40} \textbf{99.27\%} &   0.20\% &  0.29\% & 0.00\% \\
 & \textit{IPv4}  &  \cellcolor{yellow!40} \textbf{27.12\%} &  \cellcolor{yellow!40} \textbf{71.66\%} &   0.57\% &  0.32\% &   0.33\% \\
 \multirow{-2}{*}{\textbf{Quad9}} & \textit{IPv6} &  \cellcolor{yellow!40} \textbf{25.78\%} &  \cellcolor{yellow!40} \textbf{73.73\%} &   0.20\% &  0.29\% & 0.00\% \\
\rowcolor{gray!25} & \textit{IPv4} &   1.84\% &  \cellcolor{orange!40} \textbf{93.55\%} &   2.27\% &  1.25\% &   1.09\% \\
\rowcolor{gray!25} \multirow{-2}{*}{\textbf{UncensoredDNS}} & \textit{IPv6} &   0.37\% & \cellcolor{orange!40} \textbf{98.95\%} &   0.28\% &  0.40\% & 0.00\%\\
 & \textit{IPv4} &   0.55\% &   1.02\% &  \cellcolor{cyan!40} \textbf{97.68\%} &  0.47\% &   0.27\% \\
 \multirow{-2}{*}{\textbf{Yandex}}  & \textit{IPv6}  &   0.23\% &   0.18\% &  \cellcolor{cyan!40} \textbf{99.30\%} &  0.29\% & 0.00\% \\ \hline
\rowcolor{gray!25}           & \textit{IPv4} &  27.41\% &  35.66\% &  36.24\% &  0.36\% &   0.33\% \\
\rowcolor{gray!25} \multirow{-2}{*}{\textbf{Overall}} & \textit{IPv6}           &  26.04\% &  39.39\% &  34.27\% &  0.30\% & 0.00\% \\
\bottomrule

\end{tabular}
}
\end{center}

\label{table:corefrontendbuffersizes}
\end{table}

\subsection{Evaluation from the core}

\makeatletter
\newcolumntype{e}[1]{%--- Enumerated cells ---
   >{\minipage[t]{\linewidth}%
     \NoHyper%                Hyperref adds a vertical space
     \let\\\tabularnewline
    %\enumerate
      %  \addtolength{\rightskip}{0pt plus 50pt}% for raggedright
       % \setlength{\itemsep}{-\parsep}}%
       }
   p{#1}%
   <{\@finalstrut\@arstrutbox%\endenumerate
     \endNoHyper
     \endminipage}}

\newcolumntype{i}[1]{%--- Itemized cells ---
   >{\minipage[t]{\linewidth}%
        \let\\\tabularnewline
        %\itemize
           %\addtolength{\rightskip}{0pt plus 50pt}%
           %\setlength{\itemsep}{-\parsep}}%
           }
   p{#1}%
   <{\@finalstrut\@arstrutbox\endminipage}}

\begin{table*}[!t]
\caption{\em Distribution of the resolvers communicating with the authoritative name servers for uncached domains used by the public resolvers over AS.} %The ten ASs that were used in most requests over each IP version are presented individually, all others are summarized in the column \textit{'Other'}.}

\scalebox{0.8}
{
\begin{tabular}[t]{e{1.9cm}e{0.65cm}e{1.4cm}e{1.1cm}e{1.1cm}e{1.1cm}e{1.4cm}e{1.1cm}e{1.1cm}e{1.4cm}e{1.1cm}e{1.1cm}e{1cm}e{1.1cm}}\toprule

%\begin{table}[H]
%\footnotesize
%\begin{center}
%\begin{tabular}{lrrrrrrrrrrrrr}
%\toprule

&  &  Choopa &  Cogent & Packet &  Cloudfl. &  Google &  Neust. &  O-DNS &  Myth. &  W-net-1 &  FSKNET &  Yandex &  Other \\
%\hline
\midrule
\rowcolor{gray!25} & \textit{IPv4}  & \cellcolor{magenta!40} \textbf{66.58\%} & \cellcolor{magenta!40} \textbf{28.54\%} & \cellcolor{magenta!40} - & 0.64\% &  0.45\% & - & 0.05\% & - & 0.05\% & - & - & 3.69\% \\

\rowcolor{gray!25} \multirow{-2}{*}{\textbf{CleanBrowsing}} & \textit{IPv6} &         \cellcolor{magenta!40}          \textbf{59.07\%} &\cellcolor{magenta!40}  \textbf{14.89\%} &     \cellcolor{magenta!40}                 \textbf{25.26\%} &                             0.07\% &                0.21\% &                     - &                         - &                     - &                  - &                                              - &     - & 0.49\% \\

 & \textit{IPv4}& 0.06\% & 0.04\% & - & \cellcolor{orange!40} \textbf{98.04\%} & 0.46\% & - & 0.04\% & - & 0.05\% & - & - &  1.32\% \\
 
\multirow{-2}{*}{\textbf{Cloudflare}}  & \textit{IPv6}  &                    - & - &                       - &                            \cellcolor{orange!40} \textbf{99.38\%} &                0.21\% &                    - &                         -&                     - &                  - &                                              - &     - &  0.41\% \\

\rowcolor{gray!25} & \textit{IPv4}   &                \cellcolor{magenta!40}   \textbf{95.63\%} &                               0.04\% & - &                             0.67\% &                0.48\% &                     - &                         0.05\% &                     - &                  0.06\% &                                              - &     - &  3.07\% \\

\rowcolor{gray!25} \multirow{-2}{*}{\textbf{Comodo}} & \textit{IPv6}  &            \cellcolor{magenta!40}        - & - &                       - &                            - &               - &                     - &                         - &                     - &                  - & - & - &  - \\
 & \textit{IPv4}   &                    0.06\% &                               0.04\% & - &                              0.64\% &              \cellcolor{yellow!40} \textbf{97.86\%} &                     - &                         0.05\% &                     - &                  0.09\% &                                              - &     - &  1.28\% \\
 \multirow{-2}{*}{\textbf{Google}} & \textit{IPv6}  & - & - &                 - &                             0.06\% &             \cellcolor{yellow!40}  \textbf{99.41\%} &                     - &                         - &                     - &                  - &                                              - &     - &  0.52\% \\

\rowcolor{gray!25} & \textit{IPv4}   &                    0.06\% &                               0.04\% & - &                             0.63\% &                0.48\% &                    \cellcolor{green!40} \textbf{94.79\%} &                         0.05\% &                     - &                  0.06\% &                                              - &     - &  3.90\% \\

\rowcolor{gray!25} \multirow{-2}{*}{\textbf{Neustar}} & \textit{IPv6} &                    - & - & - & 0.06\% & 0.23\% & \cellcolor{green!40} \textbf{99.23\%} & - & - & - & - & -  &  0.48\% \\

& \textit{IPv4}    & 0.06\% & 0.04\% & - & 0.59\% & 0.53\% & - &                     \cellcolor{cyan!40}   \textbf{97.68\%} & - & 0.01\% & - & - &  1.10\% \\

 \multirow{-2}{*}{\textbf{OpenDNS}} & \textit{IPv6}   & - & - &                       - & 0.07\% & 0.22\% & - & \cellcolor{cyan!40} \textbf{99.30\%} & - & - & - & - &  0.41\% \\

\rowcolor{gray!25}     & \textit{IPv4} &               \cellcolor{magenta!40}    \textbf{87.92\%} &                               0.03\% & - & 0.59\% & 0.45\% & - & 0.05\% & \cellcolor{magenta!40} \textbf{9.77\%} & 0.05\% & - & - & 1.12\% \\
     
\rowcolor{gray!25} \multirow{-2}{*}{\textbf{OpenNIC}} & \textit{IPv6}  &             \cellcolor{magenta!40}      \textbf{65.21\%} & - & - & 0.06\% & 0.23\% & - & - & \cellcolor{magenta!40} \textbf{34.09\%} & - & - & - &  0.41\% \\

  & \textit{IPv4} & 0.06\% & 0.04\% & - & 0.70\% & 0.51\% & - & 0.05\% &                     - & \cellcolor{orange!40} \textbf{82.34\%} & - & - &\cellcolor{orange!40} \textbf{16.30\%} \\

 \multirow{-2}{*}{\textbf{Quad9}} & \textit{IPv6}   & - & - & - & 0.08\% &                0.22\% & - & - & - & \cellcolor{orange!40} \textbf{75.57\%} & - & - & \cellcolor{orange!40} \textbf{24.13\%} \\

\rowcolor{gray!25} & \textit{IPv4} & 0.15\% & 0.09\% & - & 1.49\% & 1.14\% &                  - & 0.12\% & - & 0.14\% & \cellcolor{yellow!40} \textbf{94.07\%} & - &  2.81\% \\
 
\rowcolor{gray!25} \multirow{-2}{*}{\textbf{UncensoredDNS}} & \textit{IPv6} &                    - & - & - & 0.08\% & 0.34\% & - & - & - & - & \cellcolor{yellow!40} \textbf{98.98\%} & - &  0.60\% \\

& \textit{IPv4}   & 0.06\% & 0.04\% & - & 0.69\% & 0.56\% & - & 0.04\% & - &                  0.05\% & - & \cellcolor{green!40} \textbf{97.43\%} &  1.13\% \\
 \multirow{-2}{*}{\textbf{Yandex}}  & \textit{IPv6} & - & - & - & 0.06\% & 0.22\% & - & - & - & - & - & \cellcolor{green!40} \textbf{99.30\%} &  0.41\% \\ %\hline

\rowcolor{gray!25} & \textit{IPv4} & 18.87\% & 3.52\% & - & 12.35\% & 12.22\% & 10.98\% & 11.83\% & 1.19\% & 9.19\% & 4.54\% & 11.75\% & 3.57\% \\

\rowcolor{gray!25} \multirow{-2}{*}{\textbf{Overall}} & \textit{IPv6}              & 14.49\% & 1.72\% & 2.95\% & 11.50\% & 11.70\% & 11.44\% & 11.51\% & 3.97\% &                  8.33\% & 7.85\% & 11.46\% &  3.09\% \\
%\hline
\bottomrule
\end{tabular}
}
%\end{center}

\label{table2:Distribution_backend}
\end{table*}

\subsubsection{\textbf{Failure rate}}
\label{failure rate from the core}

Figure \ref{fig:Image7} exhibits that the failure rates for DoTCP requests over IPv4 from the core using public resolvers are higher in the current measurement series (9.09\%) compared to previous measurements (4.01\%). CleanBrowsing, Cloudflare, Google, OpenDNS, OpenNIC, and Yandex demonstrate high resiliency with DoTCP failure rates ranging from 1.34\% to 2.62\%. Similarly, Quad9 shows higher failure rates from multiple regions and ASs. When using IPv6, the overall failure rates for all public resolvers slightly increase to 11.53\%. More READ-errors (19.67\%) and fewer Timeouts (29.99\%) are observed compared to measurements from the edge. Quad9 and UncensoredDNS exhibit the highest failure percentages. Figure \ref{fig:Image8} shows that from all continents, the DoTCP failure rates for the public resolvers is surged in comparison to that of explained in Figure \ref{fig:Image5}. However, more than 88\% of the measurements to the public resolvers over both IP versions and transport protocols receive a valid DNS response. This shows that there are no significant problems with our NSes and yields enough data to reliably analyze the resolvers’ DoTCP usage and EDNS(0).

\subsubsection{\textbf{EDNS(0) Configurations}}

To examine the usage of EDNS(0) configurations by resolvers communicating with authoritative NSes for uncached domains, we present a distribution overview in Table \ref{table2:Distribution_backend}. Most resolvers exhibit a preferred AS, with Cloudflare, Google, Neustar, OpenDNS, and Yandex DNS primarily using their own ASs for over 94\% of resolutions. Public resolvers universally employ DoUDP, emphasizing the importance of proper EDNS(0) buffer sizes in the core. Commonly used buffer sizes include 1400, 1410, and 1452 bytes. Additional buffer size details are displayed in Table \ref{table:COREbackendbuffersizes}. %Each resolver consistently advertises a preferred buffer size to NSes in over 90\% of cases, ensuring efficient communication.

\makeatletter
\newcolumntype{e}[1]{%--- Enumerated cells ---
   >{\minipage[t]{\linewidth}%
     \NoHyper%                Hyperref adds a vertical space
     \let\\\tabularnewline
    %\enumerate
      %  \addtolength{\rightskip}{0pt plus 50pt}% for raggedright
       % \setlength{\itemsep}{-\parsep}}%
       }
   p{#1}%
   <{\@finalstrut\@arstrutbox%\endenumerate
     \endNoHyper
     \endminipage}}

\newcolumntype{i}[1]{%--- Itemized cells ---
   >{\minipage[t]{\linewidth}%
        \let\\\tabularnewline
        %\itemize
           %\addtolength{\rightskip}{0pt plus 50pt}%
           %\setlength{\itemsep}{-\parsep}}%
           }
   p{#1}%
   <{\@finalstrut\@arstrutbox\endminipage}}

\begin{table}[!t]
\caption{\em EDNS(0) buffer sizes announced to the authoritative NSes. Other buffer sizes and cases in EDNS that are not used are summarized in the column "other". NOTE: CB= CleanBrowsing;U-DNS= UncensoredDNS. All the values are in percentage (\%).}

\scalebox{1.03}
{
\begin{tabular}[t]{e{0.95cm}e{0.45cm}e{0.45cm}e{0.45cm}e{0.45cm}e{0.45cm}e{0.45cm}e{0.45cm}e{0.5cm}}

%\begin{table}
%\begin{center}
%\begin{tabular}{lrrrrrrrr}
\toprule
%\hline
 & &  \textbf{512.0} &  \textbf{1232.0} &  \textbf{1400.0} &  \textbf{1410.0} &  \textbf{1452.0} &  \textbf{4096.0} &  \textbf{other} \\ 
 \hline
%\midrule
\rowcolor{gray!25} & \textit{IPv4}  &   0.11  &   \cellcolor{magenta!40} \textbf{98.24 } &    0.45  &    0.05  &    0.64  &    0.36  &   0.16  \\
\rowcolor{gray!25} \multirow{-2}{*}{\textbf{CB}} & \textit{IPv6} & 0.01  &  \cellcolor{magenta!40} \textbf{99.47 } &    0.21  &    0.00  &    0.07  &    0.05  &   0.19  \\
 & \textit{IPv4} &   0.36  &    0.65  &    0.46  &    0.04  &  \cellcolor{green!40} \textbf{98.04 } &    0.30  &   0.16  \\
\multirow{-2}{*}{\textbf{Cloudflare}}  & \textit{IPv6}    &   0.01  &    0.26  &    0.21  &    0.00  &   \cellcolor{green!40} \textbf{99.38 } &    0.04  &   0.10  \\
\rowcolor{gray!25} & \textit{IPv4}  &   0.11  &    0.70  &    0.48  &    0.05  &    0.67  &   \cellcolor{cyan!40} \textbf{95.21 } &   2.78  \\
\rowcolor{gray!25} \multirow{-2}{*}{\textbf{Comodo}} & \textit{IPv6}    &   - &  - &  - &    - &   - &   \cellcolor{cyan!40} - &   - \\
 & \textit{IPv4}  &   0.22  &    0.78  &   \cellcolor{orange!40} \textbf{97.86 } &    0.05  &    0.64  &    0.27  &   0.19  \\
\multirow{-2}{*}{\textbf{Google}} & \textit{IPv6}    &   0.02  &    0.26  &  \cellcolor{orange!40} \textbf{99.41 } &    0.00  &    0.06  &    0.14  &   0.10  \\
\rowcolor{gray!25} & \textit{IPv4}   &   0.04  &    0.70  &    0.48  &    0.05  &    0.63  &   \cellcolor{cyan!40} \textbf{97.45 } &   0.66  \\
\rowcolor{gray!25} \multirow{-2}{*}{\textbf{Neustar}} & \textit{IPv6} &   0.02  &    0.31  &    0.23  &    0.00  &    0.06  &   \cellcolor{cyan!40} \textbf{98.79 } &   0.60  \\
 & \textit{IPv4}     &   0.08  &    0.61  &    0.53  &   \cellcolor{yellow!40} \textbf{97.68 } &    0.59  &    0.32  &   0.19  \\
 \multirow{-2}{*}{\textbf{OpenDNS}}      & \textit{IPv6}   &   0.01  &    0.26  &    0.22  &   \cellcolor{yellow!40} \textbf{99.30 } &    0.07  &    0.04  &   0.10  \\
\rowcolor{gray!25}     & \textit{IPv4} &   0.06  &   \cellcolor{magenta!40} \textbf{98.29 } &    0.45  &    0.05  &    0.59  &    0.37  &   0.18  \\
\rowcolor{gray!25} \multirow{-2}{*}{\textbf{OpenNIC}} & \textit{IPv6}  &   0.01  &   \cellcolor{magenta!40} \textbf{99.56 } &    0.23  &    0.00  &    0.06  &    0.05  &   0.10  \\
 & \textit{IPv4} &   0.07  &   \cellcolor{magenta!40} \textbf{98.05 } &    0.51  &    0.05  &    0.70  &    0.40  &   0.21  \\
 \multirow{-2}{*}{\textbf{Quad9}} & \textit{IPv6}  &   0.01  &  \cellcolor{magenta!40} \textbf{99.54 } &    0.22  &    0.00  &    0.08  &    0.04  &   0.11  \\
\rowcolor{gray!25} & \textit{IPv4} &   2.68  &  \cellcolor{magenta!40} \textbf{93.15 } &    1.14  &    0.12  &    1.49  &    0.97  &   0.45  \\
\rowcolor{gray!25} \multirow{-2}{*}{\textbf{U-DNS}} & \textit{IPv6} &   1.46  &  \cellcolor{magenta!40} \textbf{97.91 } &    0.34  &    0.00  &    0.08  &    0.07  &   0.15  \\
 & \textit{IPv4} &   0.03  &    0.65  &    0.56  &    0.04  &    0.69  &   \cellcolor{cyan!40} \textbf{92.86 } &   5.16  \\
 \multirow{-2}{*}{\textbf{Yandex}}  & \textit{IPv6} &   0.00  &    0.26  &    0.22  &    0.00  &    0.06  &   \cellcolor{cyan!40} \textbf{94.31 } &   5.14  \\ %\hline
\rowcolor{gray!25}           & \textit{IPv4}   &   0.24  &   39.74  &   12.22  &   11.83  &   12.34  &   22.78  &   0.85  \\
\rowcolor{gray!25} \multirow{-2}{*}{\textbf{Overall}} & \textit{IPv6}              &   0.13  &   42.09  &   11.70  &   11.51  &   11.49  &  22.33  &   0.75  \\

%\hline
\bottomrule
\end{tabular}
}
%\end{center}

\label{table:COREbackendbuffersizes}
\end{table}

\makeatletter
\newcolumntype{e}[1]{%--- Enumerated cells ---
   >{\minipage[t]{\linewidth}%
     \NoHyper%                Hyperref adds a vertical space
     \let\\\tabularnewline
    %\enumerate
      %  \addtolength{\rightskip}{0pt plus 50pt}% for raggedright
       % \setlength{\itemsep}{-\parsep}}%
       }
   p{#1}%
   <{\@finalstrut\@arstrutbox%\endenumerate
     \endNoHyper
     \endminipage}}

\newcolumntype{i}[1]{%--- Itemized cells ---
   >{\minipage[t]{\columnwidth}%
        \let\\\tabularnewline
        %\itemize
           %\addtolength{\rightskip}{0pt plus 50pt}%
           %\setlength{\itemsep}{-\parsep}}%
           }
   p{#1}%
   <{\@finalstrut\@arstrutbox\endminipage}}

\begin{table}[!t]
\caption{\em EDNS options announced to the authoritative name servers}

\scalebox{1.17}
{
\begin{tabular}[t]{e{1.9cm}e{0.55cm}e{0.95cm}e{0.95cm}e{0.95cm}}

%\begin{table}
%\begin{center}
%\begin{tabular}{lrrrr}
%\hline
\toprule
 &  &    \textbf{EDNS} &   \textbf{Cookie} &   \textbf{ECS} \\

 %\hline
 
\midrule
\rowcolor{gray!25} & \textit{IPv4} & 99.93\% &   0.22\% &   0.10\%  \\      
\rowcolor{gray!25} \multirow{-2}{*}{\textbf{CleanBrowsing}} & \textit{IPv6} &  99.91\% &   0.05\% &   0.04\% \\
 & \textit{IPv4}  & 99.94\% &   0.32\% &   0.10\%  \\
\multirow{-2}{*}{\textbf{Cloudflare}}  & \textit{IPv6} & 100.00\% &   0.05\% &   0.05\%  \\
\rowcolor{gray!25} & \textit{IPv4}   & 98.10\% &   0.33\% &   0.11\%  \\
\rowcolor{gray!25} \multirow{-2}{*}{\textbf{Comodo}}  & \textit{IPv6}  & - &  - &  -  \\
 & \textit{IPv4}  & 99.93\% &   0.31\% &  \cellcolor{cyan!40} \textbf{14.23\%} \\
 \multirow{-2}{*}{\textbf{Google}}  & \textit{IPv6}  & 100.00\% &   0.16\% & \cellcolor{cyan!40} \textbf{12.53\%} \\
\rowcolor{gray!25} & \textit{IPv4}  & 99.93\% &   0.23\% &   0.10\%  \\
\rowcolor{gray!25} \multirow{-2}{*}{\textbf{Neustar}} & \textit{IPv6}   &  99.93\% &   0.05\% &   0.04\%  \\
 & \textit{IPv4}  & 99.94\% &   0.22\% &   0.10\% \\
 \multirow{-2}{*}{\textbf{OpenDNS}}      & \textit{IPv6}  & 100.00\% &   0.05\% &   0.04\% \\
\rowcolor{gray!25}     & \textit{IPv4}  & 99.93\% &   0.22\% &   0.11\%  \\
\rowcolor{gray!25} \multirow{-2}{*}{\textbf{OpenNIC}} & \textit{IPv6}  & 100.00\% &   0.05\% &   0.05\%  \\
 & \textit{IPv4}  & 99.93\% &   0.24\% &   0.13\%  \\
 \multirow{-2}{*}{\textbf{Quad9}} & \textit{IPv6}   & 100.00\% &   0.06\% &   0.03\%  \\
\rowcolor{gray!25} & \textit{IPv4} & 99.84\% &  \cellcolor{orange!40} \textbf{94.62\%} &   0.24\%  \\
\rowcolor{gray!25} \multirow{-2}{*}{\textbf{UncensoredDNS}} & \textit{IPv6} & 100.00\% &  \cellcolor{orange!40} \textbf{99.06\%} &   0.06\% \\
& \textit{IPv4} & 99.93\% &   0.22\% &   0.11\%  \\
\multirow{-2}{*}{\textbf{Yandex}}  & \textit{IPv6}      & 100.00\% &   0.05\% &   0.04\%  \\ \hline
\rowcolor{gray!25}           & \textit{IPv4}  & 99.93\% &   4.80\% &   1.81\%  \\
\rowcolor{gray!25} \multirow{-2}{*}{\textbf{Overall}} & \textit{IPv6}             &  99.98\% &   7.91\% &   1.49\%  \\

%\hline
\bottomrule
\end{tabular}
}
%\end{center}

\label{table:COREoptions}
\end{table}

\subsubsection{\textbf{EDNS Options}}
\label{EDNS Options}

In Table \ref{table:COREoptions}, a comprehensive listing of options employed by DNS resolvers in communication with name servers is provided. %Usage rates between IPv4 and IPv6 exhibit slight variations. 
EDNS(0) is utilized by all DNS resolvers in the majority of cases ($>$99.84\%). Among the advertised options by public resolvers, Cookie (4.80\% IPv4, 7.91\% IPv6) and EDNS Client Subnet (ECS) (1.81\% IPv4, 1.49\% IPv6) are notable but Google predominantly employs ECS (14.24\% IPv4, 12.53\% IPv6). However, other options including client subnet information, are transmitted in less than 0.24\% of requests. RFC 7871 \cite{IETF_rfc7871} specifies that NSes should include ECS with matching parameters in their response. Google indicates that if name servers do not support ECS, Google public DNS may refrain from sending ECS queries to them. This suggests that Google's usage of ECS could be higher if servers appropriately handle the requests. Nevertheless, multiple Google resolvers within the core, identifiable by their IP addresses, transmit subnet information to our servers. The question of Google's ECS usage rate when communicating with NSes that correctly respond remains open for further investigation.

\begin{table}
\caption{\em Percentage of measurements that were denoted as successful, but did not receive an answer section.}
%\tablecolorrule
\begin{center}
\begin{tabular}{lc}
\toprule
 &    \textbf{Answer Received} \\
\midrule
\textbf{CleanBrowsing}  & \cellcolor{cyan!40} \textbf{1.55\%} \\
\textbf{Cloudflare}     & \cellcolor{cyan!40} \textbf{1.63\%} \\
\textbf{Comodo}         & \cellcolor{orange!40} \textbf{62.79\%} \\
\textbf{Google}         & \cellcolor{orange!40} \textbf{57.17\%} \\
\textbf{Neustar}        & \cellcolor{orange!40} \textbf{61.37\%} \\
\textbf{OpenDNS}        & 97.65\% \\
\textbf{OpenNIC}        & \cellcolor{orange!40} \textbf{43.39\%} \\
\textbf{Quad9}          & \cellcolor{orange!40} \textbf{43.02\%} \\
\textbf{UncensoredDNS}  & \cellcolor{cyan!40} \textbf{4.05\%} \\
\textbf{Yandex}         & \cellcolor{orange!40} \textbf{64.44\%} \\ \hline
\textbf{Total} & 46.48\% \\ \hline
\end{tabular}
\end{center}

\label{table:missinganswers}
\end{table}

\begin{table}
\caption{\em Failure Rates of the RIPE Atlas measurements from the core when the unique domain generated for the request was never requested. NR = Never Requested}
\begin{center}
\begin{tabular}{lcrrr}
\toprule
&  & \textbf{Failure} &  \textbf{NR, Failure}  & \textbf{NR, Success} \\
\midrule
& \textit{IPv4} &    1.07\% &                    2.13\% & 0.42\% \\
\multirow{-2}{*}{\textbf{CleanBrowsing}} & \textit{IPv6} & 1.56\% &                    6.53\% & 0.56\%  \\
\rowcolor{gray!25} & \textit{IPv4}   &    0.56\% &                    2.16\% &        0.86\% \\
\rowcolor{gray!25} \multirow{-2}{*}{\textbf{Cloudflare}} & \textit{IPv6} &    1.42\% &                    6.20\%  & 0.81\%  \\
 \cellcolor{white} & \textit{IPv4}   &  \cellcolor{magenta!40} \textbf{24.25\%} &                    5.11\%  & 1.27\% \\
 \cellcolor{white} \multirow{-2}{*}{\textbf{Comodo}} & \textit{IPv6} &   - &                   - &  -  \\
 \rowcolor{gray!25}  & \textit{IPv4}   &    2.92\% &                    1.95\% &               0.91\%  \\
\rowcolor{gray!25} \multirow{-2}{*}{\textbf{Google}} & \textit{IPv6} & \cellcolor{magenta!40} \textbf{29.74\%} &                    6.74\% &               0.83\%  \\
 & \textit{IPv4}   &  6.19\% &                    2.69\% &     0.31\% \\
 \multirow{-2}{*}{\textbf{Neustar}} & \textit{IPv6} & \cellcolor{magenta!40} \textbf{ 33.01\%} &                    6.83\% &          0.18\% \\
\rowcolor{gray!25} & \textit{IPv4}   &    3.88\% &                    1.83\% &             0.62\% \\
\rowcolor{gray!25} \multirow{-2}{*}{\textbf{OpenDNS}} & \textit{IPv6} &  \cellcolor{magenta!40} \textbf{36.61\%} &                    6.64\% & 0.46\% \\
  & \textit{IPv4}   &    0.47\% &                \cellcolor{orange!40}   \textbf{97.31\%}  & 0.16\% \\
 \multirow{-2}{*}{\textbf{OpenNIC}} & \textit{IPv6} &    0.06\% &                   \cellcolor{orange!40} \textbf{99.31\%} & 0.01\% \\
\rowcolor{gray!25}  & \textit{IPv4}   &  \cellcolor{cyan!40}  \textbf{7.53\%} &                \cellcolor{cyan!40}    \textbf{8.91\%} &  0.69\% \\
\rowcolor{gray!25} \multirow{-2}{*}{\textbf{Quad9}} & \textit{IPv6} &  \cellcolor{magenta!40} \textbf{20.71\%} &                    6.69\% & 0.69\% \\
 & \textit{IPv4}   & 0.60\% &                  \cellcolor{orange!40} \textbf{85.85\%} & 0.19\% \\
 \multirow{-2}{*}{\textbf{UncensoredDNS}} & \textit{IPv6} &    1.18\% &                 \cellcolor{orange!40}  \textbf{10.07\%} &   1.07\%\\
\rowcolor{gray!25} & \textit{IPv4}   &    4.47\% &                    3.78\% & 0.16\%  \\
\rowcolor{gray!25} \multirow{-2}{*}{\textbf{Yandex}} & \textit{IPv6} &    5.14\% &                    6.75\% & 0.11\% \\
\hline
& \textit{IPv4}   &    5.19\% &                   21.16\% & 0.56\% \\
\multirow{-2}{*}{\textbf{All}} & \textit{IPv6} &   14.40\% &                   17.27\% &  0.52\% \\
\bottomrule
\end{tabular}
\end{center}

\label{table:failuresseries2}
\end{table}

\subsubsection{\textbf{Valid/ Invalid responses}}
\label{Valid response}

As it was observed that the transport protocol used by the probes does not affect the usage of DoTCP/ DoUDP in the communication between resolvers and the NSes, all measurements in this experiment are carried out over DoUDP. %Overall 4,576,757 individual measurements are conducted with distinct domains, 3,241,545 over IPv4, 1,335,212 over IPv6. Combining RIPE Atlas measurements and the incoming requests at the NSes yields 12,328,029 individual results. Filtering these out based on unique domain names leaves us with 11,637,539 results.
Overall, 11,637,539 individual measurements are conducted based on unique domain names. We furthermore observe that the NS returning 2KB responses receives more requests (5,642,439) than the one returning 4KB (2,395,455). Moreover, some domains are requested on both NSes (2,733,540 results). It is important to note that in many cases where a successful DNS response was indicated, the response did not contain an answer section (in 53.52\% of cases) as presented in Table \ref{table:missinganswers}. Among these cases, the vast majority (97.35\%) were truncated responses (TC-bit set), while 2.22\% denoted a server failure. %Interestingly, the latter cases were not considered as errors in RIPE Atlas. 
For the remaining 0.43\%, no clear reason for the missing answer section could be identified. The truncation of responses can be attributed to the limitation imposed by RIPE Atlas on UDP buffer sizes, which is set to a maximum of 4096 bytes which explains the high amount of truncated responses in general. % but however does not explain the big differences in the number of received answer sections between the Public resolvers presented in Table \ref{table:missinganswers}. Responses slightly larger than 4KB cannot be fully transferred, resulting in the high occurrence of truncated responses. However, this does not explain the significant variation in the number of received answer sections among the public resolvers. 
OpenDNS consistently provided a valid answer section in most cases (97.65\%), while others included an answer section in less than 64.44\% of their responses. %CleanBrowsing, Cloudflare, and Yandex predominantly sent truncated responses.
Surprisingly, RIPE Atlas measurements ended successfully even after receiving a response with the TC-bit set, indicating a lack of proper fallback to DoTCP in many probes. In Table \ref{table:failuresseries2} these cases are taken into consideration and the respective failure rates of the RIPE Atlas measurements are presented. Additionally, there were cases where certain domains were never requested at any of the servers, contributing to a failed DNS response (21.16\% IPv4, 17.27\% IPv6). %This was primarily observed with OpenNIC (97.31\% IPv4, 99.31\% IPv6) and UncensoredDNS (especially over IPv4, 85.85\%; IPv6, 10.07\%). 
Other resolvers had a failure rate of over 6.53\% for IPv6 requests that were not forwarded to authoritative name servers. This behavior may be attributed to some resolvers blacklisting our authoritative name server due to the receipt of large responses.

\subsubsection{\textbf{Canonical/ Non-canonical requests}}

We begin our analysis by classifying incoming requests as canonical and non-canonical according to Mao \emph{et al.}'s work \cite{DOT_Wild_Mao}. We then evaluate the DoTCP usage rates of the resolvers to assess their response to large buffer sizes. Additionally, we introduce a scenario involving a single incoming UDP request %(which was not part of Mao et al.'s experiment where the TC-bit was manually set). 
To focus on the resolvers' reactions to response sizes of 2KB and 4KB, we only consider results that can be directly matched to one of the servers and their respective response sizes. %We distinguish between requests answered by the 2KB and 4KB name servers. 
Table \ref{IncomingSequence_2KB4KB} displays the resolvers' usage of different scenarios when communicating with the 2KB name server. Notably, CleanBrowsing, Cloudflare, Google, OpenDNS, and UncensoredDNS predominantly send a UDP message followed by a TCP message. As indicated in Table \ref{table:COREbackendbuffersizes}, the resolvers advertise EDNS(0) buffer sizes of 1452B or less, demonstrating the expected fallback behavior. Table \ref{IncomingSequence_2KB4KB} presents the usage of different scenarios by the resolvers in response to 4KB name server replies. %Cloudflare exhibits a significantly higher number of non-canonical sequences. 
Quad9 demonstrates more non-canonical responses to 4KB compared to 2KB responses. %It is worth noting that Table \ref{table2:Distribution_backend} revealed OpenNIC's greater usage of resolvers communicating with the Core from the AS Mythic over IPv6 than IPv4, which may explain the resolvers' varying behavior across the two protocol versions.

\subsubsection{\textbf{TCP Usage}}
\label{TCP usage}

To assess TCP usage, we examined the presence of DoTCP requests within the query sequence reaching the name servers. Table \ref{TCPUsage} shows the DoTCP usage rates of resolvers when receiving 2KB responses. Resolvers primarily employing canonical scenarios consistently utilize TCP in their final request, including Quad9 (99.69\% IPv4, 99.70\% IPv6). %Neustar and OpenNIC exhibit lower TCP usage rates, with Neustar displaying significant differences between IPv4 (73.52\%) and IPv6 (72.17\%). 
Yandex and Comodo rarely use DoTCP with 2KB responses in the last request. When receiving 4KB responses (see Table \ref{TCPUsage}), almost all resolvers employ TCP in the majority of measurements for both IP versions ($>$98.67\%). However, a notable number of measurements lack TCP usage by several resolvers (up to 1.33\%), indicating possible fragmentation between the name server and resolver. %Over IPv6, OpenNIC uses TCP in less than half of the sequences reaching the name server (45.09\%). Cloudflare (95.76\% IPv4, 92.63\% IPv6) and OpenNIC (94.72\% IPv4, 41.72\% IPv6) also exhibit lower DoTCP usage rates in their final requests compared to other resolvers.

%%

% \usepackage{booktabs}
% \usepackage{multirow}
% \usepackage{graphicx}
\begin{table}[]
\centering
\caption{\em Classification of incoming sequences of DNS queries at the 2KB and 4KB name server for each resolver.}
\label{IncomingSequence_2KB4KB}
\resizebox{\columnwidth}{!}{%
\begin{tabular}{@{}ll|lll|lll@{}}
\toprule
\rowcolor[HTML]{FFFFFF} 
{\color[HTML]{333333} \textbf{Resolvers}}                                               & {\color[HTML]{333333} \textbf{}} & {\color[HTML]{333333} \textbf{}}                       & {\color[HTML]{333333} \textbf{2KB}}                                                      & {\color[HTML]{333333} \textbf{}}           & {\color[HTML]{333333} \textbf{}}                       & {\color[HTML]{333333} \textbf{4KB}}                                                      & {\color[HTML]{333333} \textbf{}}                       \\ \midrule
\rowcolor[HTML]{FFFFFF} 
{\color[HTML]{333333} }                                                                 & {\color[HTML]{333333} \textbf{}} & {\color[HTML]{333333} \textbf{Canonical}}              & {\color[HTML]{333333} \textbf{\begin{tabular}[c]{@{}l@{}}Non-\\ Canonical\end{tabular}}} & {\color[HTML]{333333} \textbf{Single UDP}} & {\color[HTML]{333333} \textbf{Canonical}}              & {\color[HTML]{333333} \textbf{\begin{tabular}[c]{@{}l@{}}Non-\\ Canonical\end{tabular}}} & {\color[HTML]{333333} \textbf{Single UDP}}             \\
\rowcolor[HTML]{EFEFEF} 
\cellcolor[HTML]{EFEFEF}{\color[HTML]{333333} }                                         & {\color[HTML]{333333} IPv4}      & \cellcolor[HTML]{FFCE93}{\color[HTML]{333333} 99.42\%} & {\color[HTML]{333333} 0.47\%}                                                            & {\color[HTML]{333333} 0.11\%}              & {\color[HTML]{333333} 98.65\%}                         & {\color[HTML]{333333} 1.27\%}                                                            & {\color[HTML]{333333} 0.08\%}                          \\
\rowcolor[HTML]{EFEFEF} 
\multirow{-2}{*}{\cellcolor[HTML]{EFEFEF}{\color[HTML]{333333} \textbf{CleanBrowsing}}} & {\color[HTML]{333333} IPv6}      & \cellcolor[HTML]{FFCE93}{\color[HTML]{333333} 99.56\%} & {\color[HTML]{333333} 0.24\%}                                                            & {\color[HTML]{333333} 0.21\%}              & {\color[HTML]{333333} 98.83\%}                         & {\color[HTML]{333333} 0.25\%}                                                            & {\color[HTML]{333333} 0.92\%}                          \\
\rowcolor[HTML]{FFFFFF} 
\cellcolor[HTML]{FFFFFF}{\color[HTML]{333333} }                                         & {\color[HTML]{333333} IPv4}      & \cellcolor[HTML]{FFCE93}{\color[HTML]{333333} 95.36\%} & {\color[HTML]{333333} 4.45\%}                                                            & {\color[HTML]{333333} 0.19\%}              & \cellcolor[HTML]{FFCE93}{\color[HTML]{333333} 91.76\%} & {\color[HTML]{333333} 8.22\%}                                                            & {\color[HTML]{333333} 0.00\%}                          \\
\rowcolor[HTML]{FFFFFF} 
\multirow{-2}{*}{\cellcolor[HTML]{FFFFFF}{\color[HTML]{333333} \textbf{Cloudflare}}}    & {\color[HTML]{333333} IPv6}      & \cellcolor[HTML]{FFCE93}{\color[HTML]{333333} 91.96\%} & {\color[HTML]{333333} 7.70\%}                                                            & {\color[HTML]{333333} 0.34\%}              & \cellcolor[HTML]{FFCE93}{\color[HTML]{333333} 85.24\%} & {\color[HTML]{333333} 14.37\%}                                                           & {\color[HTML]{333333} 0.36\%}                          \\
\rowcolor[HTML]{EFEFEF} 
\cellcolor[HTML]{EFEFEF}{\color[HTML]{333333} }                                         & {\color[HTML]{333333} IPv4}      & {\color[HTML]{333333} 1.19\%}                          & \cellcolor[HTML]{FFCE93}{\color[HTML]{333333} 97.83\%}                                   & {\color[HTML]{333333} 0.98\%}              & {\color[HTML]{333333} 8.91\%}                          & {\color[HTML]{333333} 91.07\%}                                                           & {\color[HTML]{333333} 0.02\%}                          \\
\rowcolor[HTML]{EFEFEF} 
\multirow{-2}{*}{\cellcolor[HTML]{EFEFEF}{\color[HTML]{333333} \textbf{Comodo}}}        & {\color[HTML]{333333} IPv6}      & {\color[HTML]{333333} -}                               & {\color[HTML]{333333} -}                                                                 & {\color[HTML]{333333} -}                   & {\color[HTML]{333333} -}                               & {\color[HTML]{333333} -}                                                                 & {\color[HTML]{333333} -}                               \\
\rowcolor[HTML]{FFFFFF} 
\cellcolor[HTML]{FFFFFF}{\color[HTML]{333333} }                                         & {\color[HTML]{333333} IPv4}      & \cellcolor[HTML]{FFCE93}{\color[HTML]{333333} 99.24\%} & {\color[HTML]{333333} 0.67\%}                                                            & {\color[HTML]{333333} 0.10\%}              & {\color[HTML]{333333} 98.67\%}                         & {\color[HTML]{333333} 0.83\%}                                                            & {\color[HTML]{333333} 0.51\%}                          \\
\rowcolor[HTML]{FFFFFF} 
\multirow{-2}{*}{\cellcolor[HTML]{FFFFFF}{\color[HTML]{333333} \textbf{Google}}}        & {\color[HTML]{333333} IPv6}      & \cellcolor[HTML]{FFCE93}{\color[HTML]{333333} 99.44\%} & {\color[HTML]{333333} 0.26\%}                                                            & {\color[HTML]{333333} 0.30\%}              & {\color[HTML]{333333} 98.82\%}                         & {\color[HTML]{333333} 0.19\%}                                                            & {\color[HTML]{333333} 0.99\%}                          \\
\rowcolor[HTML]{EFEFEF} 
\cellcolor[HTML]{EFEFEF}{\color[HTML]{333333} }                                         & {\color[HTML]{333333} IPv4}      & {\color[HTML]{333333} 1.31\%}                          & {\color[HTML]{333333} 97.61\%}                                                           & {\color[HTML]{333333} 1.08\%}              & {\color[HTML]{333333} 3.13\%}                          & {\color[HTML]{333333} 96.86\%}                                                           & {\color[HTML]{333333} 0.01\%}                          \\
\rowcolor[HTML]{EFEFEF} 
\multirow{-2}{*}{\cellcolor[HTML]{EFEFEF}{\color[HTML]{333333} \textbf{Neustar}}}       & {\color[HTML]{333333} IPv6}      & {\color[HTML]{333333} 0.25\%}                          & {\color[HTML]{333333} 98.67\%}                                                           & {\color[HTML]{333333} 1.07\%}              & {\color[HTML]{333333} 0.73\%}                          & {\color[HTML]{333333} 98.25\%}                                                           & {\color[HTML]{333333} 1.02\%}                          \\
\rowcolor[HTML]{FFFFFF} 
\cellcolor[HTML]{FFFFFF}{\color[HTML]{333333} }                                         & {\color[HTML]{333333} IPv4}      & {\color[HTML]{333333} 97.71\%}                         & {\color[HTML]{333333} 2.08\%}                                                            & {\color[HTML]{333333} 0.21\%}              & {\color[HTML]{333333} 97.55\%}                         & {\color[HTML]{333333} 2.37\%}                                                            & {\color[HTML]{333333} 0.08\%}                          \\
\rowcolor[HTML]{FFFFFF} 
\multirow{-2}{*}{\cellcolor[HTML]{FFFFFF}{\color[HTML]{333333} \textbf{OpenDNS}}}       & {\color[HTML]{333333} IPv6}      & {\color[HTML]{333333} 99.13\%}                         & {\color[HTML]{333333} 0.62\%}                                                            & {\color[HTML]{333333} 0.25\%}              & {\color[HTML]{333333} 98.38\%}                         & {\color[HTML]{333333} 0.79\%}                                                            & {\color[HTML]{333333} 0.83\%}                          \\
\rowcolor[HTML]{EFEFEF} 
\cellcolor[HTML]{EFEFEF}{\color[HTML]{333333} }                                         & {\color[HTML]{333333} IPv4}      & {\color[HTML]{333333} 55.68\%}                         & {\color[HTML]{333333} 35.60\%}                                                           & {\color[HTML]{333333} 8.72\%}              & \cellcolor[HTML]{B6E8EC}{\color[HTML]{333333} 64.36\%} & \cellcolor[HTML]{B6E8EC}{\color[HTML]{333333} 35.42\%}                                   & \cellcolor[HTML]{B6E8EC}{\color[HTML]{333333} 0.22\%}  \\
\rowcolor[HTML]{EFEFEF} 
\multirow{-2}{*}{\cellcolor[HTML]{EFEFEF}{\color[HTML]{333333} \textbf{OpenNIC}}}       & {\color[HTML]{333333} IPv6}      & {\color[HTML]{333333} 35.20\%}                         & {\color[HTML]{333333} 26.89\%}                                                           & {\color[HTML]{333333} 37.91\%}             & \cellcolor[HTML]{B6E8EC}{\color[HTML]{333333} 38.34\%} & \cellcolor[HTML]{B6E8EC}{\color[HTML]{333333} 7.06\%}                                    & \cellcolor[HTML]{B6E8EC}{\color[HTML]{333333} 54.60\%} \\
\rowcolor[HTML]{FFFFFF} 
\cellcolor[HTML]{FFFFFF}{\color[HTML]{333333} }                                         & {\color[HTML]{333333} IPv4}      & \cellcolor[HTML]{FFFC9E}{\color[HTML]{333333} 46.14\%} & {\color[HTML]{333333} 53.67\%}                                                           & {\color[HTML]{333333} 0.19\%}              & {\color[HTML]{333333} 26.37\%}                         & {\color[HTML]{333333} 73.60\%}                                                           & {\color[HTML]{333333} 0.03\%}                          \\
\rowcolor[HTML]{FFFFFF} 
\multirow{-2}{*}{\cellcolor[HTML]{FFFFFF}{\color[HTML]{333333} \textbf{Quad9}}}         & {\color[HTML]{333333} IPv6}      & \cellcolor[HTML]{FFFC9E}{\color[HTML]{333333} 46.28\%} & {\color[HTML]{333333} 53.47\%}                                                           & {\color[HTML]{333333} 0.25\%}              & {\color[HTML]{333333} 31.23\%}                         & {\color[HTML]{333333} 68.21\%}                                                           & {\color[HTML]{333333} 0.56\%}                          \\
\rowcolor[HTML]{EFEFEF} 
\cellcolor[HTML]{EFEFEF}{\color[HTML]{333333} }                                         & {\color[HTML]{333333} IPv4}      & {\color[HTML]{333333} 94.11\%}                         & {\color[HTML]{333333} 4.70\%}                                                            & {\color[HTML]{333333} 1.19\%}              & {\color[HTML]{333333} 94.52\%}                         & {\color[HTML]{333333} 5.43\%}                                                            & {\color[HTML]{333333} 0.05\%}                          \\
\rowcolor[HTML]{EFEFEF} 
\multirow{-2}{*}{\cellcolor[HTML]{EFEFEF}{\color[HTML]{333333} \textbf{UncensoredDNS}}} & {\color[HTML]{333333} IPv6}      & {\color[HTML]{333333} 99.52\%}                         & {\color[HTML]{333333} 0.20\%}                                                            & {\color[HTML]{333333} 0.29\%}              & {\color[HTML]{333333} 99.48\%}                         & {\color[HTML]{333333} 0.11\%}                                                            & {\color[HTML]{333333} 0.41\%}                          \\
\rowcolor[HTML]{FFFFFF} 
\cellcolor[HTML]{FFFFFF}{\color[HTML]{333333} }                                         & {\color[HTML]{333333} IPv4}      & {\color[HTML]{333333} 1.19\%}                          & {\color[HTML]{333333} 97.79\%}                                                           & {\color[HTML]{333333} 1.02\%}              & {\color[HTML]{333333} 3.34\%}                          & \cellcolor[HTML]{FFCE93}{\color[HTML]{333333} 96.56\%}                                   & {\color[HTML]{333333} 0.10\%}                          \\
\rowcolor[HTML]{FFFFFF} 
\multirow{-2}{*}{\cellcolor[HTML]{FFFFFF}{\color[HTML]{333333} \textbf{Yandex}}}        & {\color[HTML]{333333} IPv6}      & {\color[HTML]{333333} 0.22\%}                          & {\color[HTML]{333333} 98.94\%}                                                           & {\color[HTML]{333333} 0.84\%}              & {\color[HTML]{333333} 0.83\%}                          & \cellcolor[HTML]{FFCE93}{\color[HTML]{333333} 97.86\%}                                   & {\color[HTML]{333333} 1.31\%}                          \\
\rowcolor[HTML]{EFEFEF} 
\cellcolor[HTML]{EFEFEF}{\color[HTML]{333333} }                                         & {\color[HTML]{333333} IPv4}      & {\color[HTML]{333333} 59.69\%}                         & {\color[HTML]{333333} 39.82\%}                                                           & {\color[HTML]{333333} 0.48\%}              & {\color[HTML]{333333} 70.91\%}                         & {\color[HTML]{333333} 28.95\%}                                                           & {\color[HTML]{333333} 0.13\%}                          \\
\rowcolor[HTML]{EFEFEF} 
\multirow{-2}{*}{\cellcolor[HTML]{EFEFEF}{\color[HTML]{333333} \textbf{All}}}           & {\color[HTML]{333333} IPv6}      & {\color[HTML]{333333} 69.05\%}                         & {\color[HTML]{333333} 30.50\%}                                                           & {\color[HTML]{333333} 0.45\%}              & {\color[HTML]{333333} 75.43\%}                         & {\color[HTML]{333333} 23.79\%}                                                           & {\color[HTML]{333333} 0.78\%}                          \\ %\cmidrule(l){2-8}

\bottomrule
\end{tabular}%
}
\end{table}

% \usepackage{booktabs}
% \usepackage{multirow}
% \usepackage{graphicx}
\begin{table}[]
\centering
\caption{\em TCP usage of DNS resolvers when 2KB and 4KB responses are received. \textit{TCP Used} represents all scenarios in which TCP is used at any point in the request sequence. For \textit{Last TCP}, only those sequences ending with a DoTCP request are considered.}
\label{TCPUsage}
\resizebox{\columnwidth}{!}{%
\begin{tabular}{@{}ll|ll|ll@{}}
\toprule
\textbf{}                           \textbf{Resolvers}            &      & \multicolumn{2}{l}{\textbf{2KB}}                                  & \multicolumn{2}{l}{\textbf{4KB}}                                  \\ \midrule
\textbf{}                                            &      & \textbf{TCP Used}               & \textbf{Last TCP}               & \textbf{TCP Used}               & \textbf{Last TCP}               \\
                                                              & IPv4 & 99.84\%                         & 99.80\%                         & 99.92\%                         & 99.77\%                         \\
\multirow{-2}{*}{\textbf{CleanBrowsing}}                      & IPv6 & 99.76\%                         & 99.76\%                         & 99.08\%                         & 99.02\%                         \\
\rowcolor[HTML]{E3E3E3} 
\cellcolor[HTML]{E3E3E3}                                      & IPv4 & 99.74\%                         & 96.95\%                         & 100.00\%                        & \cellcolor[HTML]{FFCE93}95.76\% \\
\rowcolor[HTML]{E3E3E3} 
\multirow{-2}{*}{\cellcolor[HTML]{E3E3E3}\textbf{Cloudflare}} & IPv6 & 99.60\%                         & 95.84\%                         & 99.64\%                         & \cellcolor[HTML]{FFCE93}92.63\% \\
                                                              & IPv4 & \cellcolor[HTML]{FFCE93}7.94\%  & \cellcolor[HTML]{FFCE93}3.36\%  & 99.98\%                         & 99.52\%                         \\
\multirow{-2}{*}{\textbf{Comodo}}                             & IPv6 & -                               & -                               &                                 &                                 \\
\rowcolor[HTML]{E3E3E3} 
\cellcolor[HTML]{E3E3E3}                                      & IPv4 & 99.86\%                         & 99.81\%                         & 99.49\%                         & 99.42\%                         \\
\rowcolor[HTML]{E3E3E3} 
\multirow{-2}{*}{\cellcolor[HTML]{E3E3E3}\textbf{Google}}     & IPv6 & 99.65\%                         & 99.65\%                         & 99.00\%                         & 98.97\%                         \\
                                                              & IPv4 & \cellcolor[HTML]{FFCE93}73.52\% & \cellcolor[HTML]{FFCE93}49.96\% & 99.99\%                         & 99.78\%                         \\
\multirow{-2}{*}{\textbf{Neustar}}                            & IPv6 & \cellcolor[HTML]{FFCE93}72.17\% & \cellcolor[HTML]{FFCE93}48.46\% & 98.98\%                         & 98.91\%                         \\
\rowcolor[HTML]{E3E3E3} 
\cellcolor[HTML]{E3E3E3}                                      & IPv4 & 99.73\%                         & 99.67\%                         & 99.91\%                         & 99.81\%                         \\
\rowcolor[HTML]{E3E3E3} 
\multirow{-2}{*}{\cellcolor[HTML]{E3E3E3}\textbf{OpenDNS}}    & IPv6 & 99.70\%                         & 99.70\%                         & 99.16\%                         & 99.13\%                         \\
                                                              & IPv4 & 88.05\%                         & 85.16\%                         & 99.78\%                         & 94.72\%                         \\
\multirow{-2}{*}{\textbf{OpenNIC}}                            & IPv6 & 54.35\%                         & 54.35\%                         & \cellcolor[HTML]{FFCE93}45.09\% & \cellcolor[HTML]{FFCE93}41.72\% \\
\rowcolor[HTML]{E3E3E3} 
\cellcolor[HTML]{E3E3E3}                                      & IPv4 & 99.74\%                         & 99.69\%                         & 99.97\%                         & 99.80\%                         \\
\rowcolor[HTML]{E3E3E3} 
\multirow{-2}{*}{\cellcolor[HTML]{E3E3E3}\textbf{Quad9}}      & IPv6 & 99.70\%                         & 99.70\%                         & 99.43\%                         & 99.37\%                         \\
                                                              & IPv4 & 98.34\%                         & 98.04\%                         & 99.95\%                         & 99.21\%                         \\
\multirow{-2}{*}{\textbf{UncensoredDNS}}                      & IPv6 & 99.66\%                         & 99.66\%                         & 99.59\%                         & 99.58\%                         \\
\rowcolor[HTML]{E3E3E3} 
\cellcolor[HTML]{E3E3E3}                                      & IPv4 & \cellcolor[HTML]{FFCE93}4.49\%  & \cellcolor[HTML]{FFCE93}3.17\%  & 99.85\%                         & 99.54\%                         \\
\rowcolor[HTML]{E3E3E3} 
\multirow{-2}{*}{\cellcolor[HTML]{E3E3E3}\textbf{Yandex}}     & IPv6 & \cellcolor[HTML]{FFCE93}1.58\%  & \cellcolor[HTML]{FFCE93}0.94\%  & 98.67\%                         & 98.58\%                         \\
                                                              & IPv4 & 75.36\%                         & 71.97\%                         & 99.86\%                         & 98.79\%                         \\
\multirow{-2}{*}{\textbf{All}}                                & IPv6 & 84.25\%                         & 81.38\%                         & 99.22\%                         & 97.84\%                         \\ %\cmidrule(l){2-6} 
\bottomrule
\end{tabular}%
}
\end{table}

\section{Limitations and Future Work}
\label{limitations}

Approximately 88\% of our probe measurements are concentrated in North America and Europe, limiting the generality of DNS resiliency observations to other regions. To address this limitation, we provide response times categorized by continent. However, it is important to note that observations for continents with fewer probes have smaller sample sizes, which hinders drawing reliable conclusions. Similarly, when analyzing response times of specific autonomous systems, particularly over IPv6, the sample size remains relatively low. The study of EDNS(0) options focuses on the communication between different resolvers and our custom authoritative NSes. Therefore, the usage numbers may not accurately represent the capabilities of the resolvers and their EDNS(0) options in general. The observations reveal various non-canonical sequences employed by DNS resolvers in response to large response sizes. Further investigation is required to fully understand the behavior of different resolvers, including their adjustment of announced EDNS(0) buffer sizes when receiving large responses.

While our study emphasized on the unencrypted DNS protocols DoUDP and DoTCP, the recently standardized encrypted DNS protocol DNS-over-QUIC (DoQ) (RFC 9250)\cite{IETF_rfc9250}\cite{DoQ_Vaibhav}\cite{WebPrivacy_Design_IFIP23}\cite{QUIC_CrossLayer_Jayasree} does inherently solve fragmentation by means of the QUIC protocol (RFC 9000)\cite{IETF_rfc9000} while also supporting increased DNS message sizes. However, DoQ adoption currently is scarce \cite{mike:pam:2022}; yet, DNS over QUIC is a promising candidate to supersede both DoUDP and DoTCP in the future, thereby warranting a detailed investigation when DoQ adoption rises.

\section{Conclusion}
\label{conclusion}

We conducted the measurements analyzing DoTCP resiliency, responsiveness and deployment from the edge and the core over IPv4 and IPv6. Additionally, the EDNS(0) configurations of ten public resolvers were studied. Issuing more than 14M individual DNS requests using 2527 globally distributed RIPE Atlas probes, we performed multiple experiments focusing on observations to conclude that most resolvers show similar resiliency for both DoTCP and DoUDP where 3 out of 10 resolvers mainly announce very large EDNS(0) buffer sizes, which potentially causes fragmentation. The analysis of DoTCP and DoUDP performance revealed significant regional variations for both IP versions. Notably, requests originating from Africa or South America exhibited the highest median response times. This highlights the need for further investigation and optimization in such regions. Particularly over IPv4, Cloudflare and Google emerged as the Public resolvers with the most consistent and stable response times across all continents. In reaction to large response sizes from authoritative name servers, we find that resolvers do not fall back to the usage of DoTCP in many cases, bearing the risk of fragmented responses. As the message sizes in the DNS are expected to grow further, this problem will become more urgent in the future.

\section*{Acknowledgment}

This work was supported by the Volkswagenstiftung Niedersächsisches Vorab (Funding No. ZN3695).

\bibliographystyle{unsrt}
\footnotesize
\bibliography{references}

\begin{thebibliography}{10}

\bibitem{DNSDevelopment_Paul}
Paul~V. Mockapetris and Kevin~J. Dunlap.
\newblock Development of the domain name system.
\newblock In Vinton~G. Cerf, editor, {\em Proceedings of the {ACM} Symposium on
  Communications Architectures and Protocols, {SIGCOMM} 1988, Stanford, CA,
  USA, August 16-18, 1988}, pages 123--133. {ACM}, 1988.

\bibitem{IETF_rfc1034}
Paul~V. Mockapetris.
\newblock Domain names - concepts and facilities.
\newblock {\em {RFC}}, 1034:1--55, 1987.

\bibitem{IETF_rfc793}
Jon Postel.
\newblock Transmission control protocol.
\newblock {\em {RFC}}, 793:1--91, 1981.

\bibitem{IETF_rfc768}
Jon Postel.
\newblock User datagram protocol.
\newblock {\em {RFC}}, 768:1--3, 1980.

\bibitem{ietf_RFC_1035}
Paul~V. Mockapetris.
\newblock Domain names - implementation and specification.
\newblock {\em {RFC}}, 1035:1--55, 1987.

\bibitem{IETF_rfc7766}
John Dickinson, Sara Dickinson, Ray Bellis, Allison Mankin, and Duane Wessels.
\newblock {DNS} transport over {TCP} - implementation requirements.
\newblock {\em {RFC}}, 7766:1--19, 2016.

\bibitem{IETF_rfc5966}
Ray Bellis.
\newblock {DNS} transport over {TCP} - implementation requirements.
\newblock {\em {RFC}}, 5966:1--7, 2010.

\bibitem{IETF_rfc2671}
Paul Vixie.
\newblock Extension mechanisms for {DNS} {(EDNS0)}.
\newblock {\em {RFC}}, 2671:1--7, 1999.

\bibitem{Ivica_DNS_Compliance}
Ivica Stipovic.
\newblock Analysis of an extension dynamic name service - {A} discussion on
  {DNS} compliance with {RFC} 6891.
\newblock {\em CoRR}, abs/2003.13319, 2020.

\bibitem{koolhaas_defragmentingDNS}
Axel Koolhaas and Tjeerd Slokkker.
\newblock {Defragmenting DNS: Determining the Optimal Maximum UDP Response Size
  for DNS}, 2020.
\newblock [Last Accessed: 19.April.2023]: https://bit.ly/3Ag6Mck.

\bibitem{DoTCP_Kosek}
Mike Kosek, Trinh~Viet Doan, Simon Huber, and Vaibhav Bajpai.
\newblock {Measuring {DNS} over {TCP} in the Era of increasing {DNS} Response
  Sizes: A View from the Edge}.
\newblock {\em Computer Communication Review}, 2022.

\bibitem{bajpai:ton:2019}
Vaibhav Bajpai and J{\"{u}}rgen Sch{\"{o}}nw{\"{a}}lder.
\newblock {A Longitudinal View of Dual-Stacked Websites - Failures, Latency and
  Happy Eyeballs}.
\newblock {\em {IEEE/ACM} Trans. Netw.}, 27(2):577--590, 2019.

\bibitem{Internet_resilience_IFIP23}
Pratyush Dikshit, Mike Kosek, Nils Faulhaber, Jayasree Sengupta, and Vaibhav
  Bajpai.
\newblock Evaluating dns resiliency with truncation, fragmentation and dotcp
  fallback.
\newblock In {\em IFIP Networking Conference}, 2023.

\bibitem{fragmentation_moura}
Giovane C.~M. Moura, Moritz M{\"{u}}ller, Marco Davids, Maarten Wullink, and
  Cristian Hesselman.
\newblock {Fragmentation, Truncation, and Timeouts: Are Large {DNS} Messages
  Falling to Bits?}
\newblock In {\em Passive and Active Measurement Conference (PAM)}, 2021.

\bibitem{IETF_rfc8900}
Ron Bonica, Fred Baker, Geoff Huston, Robert~M. Hinden, Ole Troan, and Fernando
  Gont.
\newblock {IP} fragmentation considered fragile.
\newblock {\em {RFC}}, 8900:1--23, 2020.

\bibitem{Fragmentation_Amir}
Amir Herzberg and Haya Shulman.
\newblock Fragmentation considered poisonous, or:
  One-domain-to-rule-them-all.org.
\newblock In {\em {IEEE} Conference on Communications and Network Security
  (CNS) 2013}. {IEEE}, 2013.

\bibitem{Fragmentation_leaking_Shulman}
Haya Shulman and Michael Waidner.
\newblock {Fragmentation Considered Leaking: Port Inference for {DNS}
  Poisoning}.
\newblock In {\em Applied Cryptography and Network Security Conference,
  (ACNS)}. Springer, 2014.

\bibitem{DNSPoisoning_Sooel}
Sooel Son and Vitaly Shmatikov.
\newblock The hitchhiker's guide to {DNS} cache poisoning.
\newblock In Sushil Jajodia and Jianying Zhou, editors, {\em Security and
  Privacy in Communication Networks - 6th Iternational {ICST} Conference,
  SecureComm 2010, Singapore, September 7-9, 2010. Proceedings}, volume~50 of
  {\em Lecture Notes of the Institute for Computer Sciences, Social Informatics
  and Telecommunications Engineering}, pages 466--483. Springer, 2010.

\bibitem{DNS_Poisoning_BergerDG21}
Harel Berger, Amit~Z. Dvir, and Moti Geva.
\newblock {A wrinkle in time: A case study in {DNS} poisoning}.
\newblock {\em International Journal of Information Security}, 20(3):313--329,
  2021.

\bibitem{CERT_DNS_poisoning}
Chad~R Dougherty.
\newblock Multiple dns implementations vulnerable to cache poisoning,
  https://www.kb.cert.org/vuls/id/800113, 2008.

\bibitem{IETF_rfc2464}
Matt Crawford.
\newblock {Transmission of IPv6 Packets over Ethernet Networks}.
\newblock {\em {RFC}}, 2464:1--7, 1998.

\bibitem{Encoding_DNS_CaoMWL17}
Jin Cao, Maode Ma, Xilei Wang, and Haochen Liu.
\newblock A selective re-query case sensitive encoding scheme against {DNS}
  cache poisoning attacks.
\newblock {\em Wireless Personal Communications}, 94(3):1263--1279, 2017.

\bibitem{DoTCP_Poisoning_Dai}
Tianxiang Dai, Haya Shulman, and Michael Waidner.
\newblock Dns-over-tcp considered vulnerable.
\newblock In {\em {ANRW} '21: Applied Networking Research Workshop, Virtual
  Event, USA, July 24-30, 2021}, pages 76--81. {ACM}, 2021.

\bibitem{weaver_dnsMeasurements}
Nicholas Weaver, Christian Kreibich, Boris Nechaev, and Vern Paxson.
\newblock Implications of netalyzr’s dns measurements.
\newblock In {\em Proceedings of the First Workshop on Securing and Trusting
  Internet Names (SATIN), Teddington, United Kingdom}, 2011.

\bibitem{DoTLS_rfc7858}
Zi~Hu, Liang Zhu, John~S. Heidemann, Allison Mankin, Duane Wessels, and Paul~E.
  Hoffman.
\newblock Specification for {DNS} over transport layer security {(TLS)}.
\newblock {\em {RFC}}, 7858:1--19, 2016.

\bibitem{DoH_rfc8484}
Paul~E. Hoffman and Patrick McManus.
\newblock {DNS} queries over {HTTPS} (doh).
\newblock {\em {RFC}}, 8484:1--21, 2018.

\bibitem{Dikshit_DNS_Privacy}
Pratyush Dikshit, Jayasree Sengupta, and Vaibhav Bajpai.
\newblock Recent trends on privacy-preserving technologies under
  standardization at the {IETF}.
\newblock {\em CoRR}, abs/2301.01124, 2023.

\bibitem{DNS_Encryption_Chaoyi}
Chaoyi Lu, Baojun Liu, Zhou Li, Shuang Hao, Hai{-}Xin Duan, Mingming Zhang,
  Chunying Leng, Ying Liu, Zaifeng Zhang, and Jianping Wu.
\newblock An end-to-end, large-scale measurement of dns-over-encryption: How
  far have we come?
\newblock In {\em Proceedings of the Internet Measurement Conference, {IMC}
  2019, Amsterdam, The Netherlands, October 21-23, 2019}, pages 22--35. {ACM},
  2019.

\bibitem{DNS_privacy_Deccio}
Casey~T. Deccio and Jacob Davis.
\newblock {DNS} privacy in practice and preparation.
\newblock In Aziz Mohaisen and Zhi{-}Li Zhang, editors, {\em Proceedings of the
  15th International Conference on Emerging Networking Experiments And
  Technologies, CoNEXT 2019, Orlando, FL, USA, December 09-12, 2019}, pages
  138--143. {ACM}, 2019.

\bibitem{DoTLS_Doan}
Trinh~Viet Doan, Irina Tsareva, and Vaibhav Bajpai.
\newblock Measuring {DNS} over {TLS} from the edge: Adoption, reliability, and
  response times.
\newblock In Oliver Hohlfeld, Andra Lutu, and Dave Levin, editors, {\em Passive
  and Active Measurement - 22nd International Conference, {PAM} 2021, Virtual
  Event, March 29 - April 1, 2021, Proceedings}, volume 12671 of {\em Lecture
  Notes in Computer Science}, pages 192--209. Springer, 2021.

\bibitem{DNS_Danzig}
Peter~B. Danzig, Katia Obraczka, and Anant Kumar.
\newblock An analysis of wide-area name server traffic: {A} study of the
  internet domain name system.
\newblock In Deepinder~P. Sidhu and David Oran, editors, {\em Proceedings of
  the Conference on Communications Architecture {\&} Protocols, {SIGCOMM} 1992,
  Baltimore, Maryland, USA, August 17-20, 1992}, pages 281--292. {ACM}, 1992.

\bibitem{DNS_Performance_Jung}
Jaeyeon Jung, Emil Sit, Hari Balakrishnan, and Robert~Tappan Morris.
\newblock {DNS} performance and the effectiveness of caching.
\newblock In Vern Paxson, editor, {\em Proceedings of the 1st {ACM} {SIGCOMM}
  Internet Measurement Workshop, {IMW} 2001, San Francisco, California, USA,
  November 1-2, 2001}, pages 153--167. {ACM}, 2001.

\bibitem{Response_time_Ager}
Bernhard Ager, Wolfgang M{\"{u}}hlbauer, Georgios Smaragdakis, and Steve Uhlig.
\newblock Comparing {DNS} resolvers in the wild.
\newblock In Mark Allman, editor, {\em Proceedings of the 10th {ACM} {SIGCOMM}
  Internet Measurement Conference, {IMC} 2010, Melbourne, Australia - November
  1-3, 2010}, pages 15--21. {ACM}, 2010.

\bibitem{DNS_Evolution_Otto}
John~S. Otto, Mario~A. S{\'{a}}nchez, John~P. Rula, and Fabi{\'{a}}n~E.
  Bustamante.
\newblock Content delivery and the natural evolution of {DNS:} remote dns
  trends, performance issues and alternative solutions.
\newblock In John~W. Byers, Jim Kurose, Ratul Mahajan, and Alex~C. Snoeren,
  editors, {\em Proceedings of the 12th {ACM} {SIGCOMM} Internet Measurement
  Conference, {IMC} '12, Boston, MA, USA, November 14-16, 2012}, pages
  523--536. {ACM}, 2012.

\bibitem{Proactive_Caching_Cohen}
Edith Cohen and Haim Kaplan.
\newblock Proactive caching of {DNS} records: addressing a performance
  bottleneck.
\newblock {\em Comput. Networks}, 41(6):707--726, 2003.

\bibitem{DNSPerformance_lookups_Park}
KyoungSoo Park, Vivek~S. Pai, Larry~L. Peterson, and Zhe Wang.
\newblock Codns: Improving {DNS} performance and reliability via cooperative
  lookups.
\newblock In Eric~A. Brewer and Peter Chen, editors, {\em 6th Symposium on
  Operating System Design and Implementation {(OSDI} 2004), San Francisco,
  California, USA, December 6-8, 2004}, pages 199--214. {USENIX} Association,
  2004.

\bibitem{PublicDNS_Doan}
Trinh~Viet Doan, Justus Fries, and Vaibhav Bajpai.
\newblock Evaluating public {DNS} services in the wake of increasing
  centralization of {DNS}.
\newblock In Zheng Yan, Gareth Tyson, and Dimitrios Koutsonikolas, editors,
  {\em {IFIP} Networking Conference, {IFIP} Networking 2021, Espoo and
  Helsinki, Finland, June 21-24, 2021}, pages 1--9. {IEEE}, 2021.

\bibitem{DNSSEC_Fragmentation_Broek_Commag}
Gijs Van Den~Broek, Roland Van Rijswijk-Deij, Anna Sperotto, and Aiko Pras.
\newblock {DNSSEC meets real world: Dealing with unreachability caused by
  fragmentation}.
\newblock {\em IEEE Communications Magazine}, 2014.

\bibitem{bajpaiccr2015}
Vaibhav Bajpai, Steffie~Jacob Eravuchira, and J{\"{u}}rgen
  Sch{\"{o}}nw{\"{a}}lder.
\newblock {Lessons Learned From Using the {RIPE} Atlas Platform for Measurement
  Research}.
\newblock {\em Computer Communications Review}, 45(3):35--42, 2015.

\bibitem{RIPE_Holterbach}
Thomas Holterbach, Cristel Pelsser, Randy Bush, and Laurent Vanbever.
\newblock Quantifying interference between measurements on the {RIPE} atlas
  platform.
\newblock In Kenjiro Cho, Kensuke Fukuda, Vivek~S. Pai, and Neil Spring,
  editors, {\em Proceedings of the 2015 {ACM} Internet Measurement Conference,
  {IMC} 2015, Tokyo, Japan, October 28-30, 2015}, pages 437--443. {ACM}, 2015.

\bibitem{bajpai:im:2017}
Vaibhav Bajpai et~al.
\newblock {Vantage point selection for IPv6 measurements: Benefits and
  limitations of {RIPE} Atlas tags}.
\newblock In {\em {IFIP/IEEE} Symposium on Integrated Network and Service
  Management (IM)}, 2017.

\bibitem{IETF_rfc7871}
Carlo Contavalli, Wilmer van~der Gaast, David~C. Lawrence, and Warren Kumari.
\newblock Client subnet in {DNS} queries.
\newblock {\em {RFC}}, 7871:1--30, 2016.

\bibitem{DOT_Wild_Mao}
Jiarun Mao, Michael Rabinovich, and Kyle Schomp.
\newblock {Assessing Support for DNS-over-TCP in the Wild}.
\newblock In {\em Passive and Active Measurement Conference (PAM)}, pages
  487--517. Springer, 2022.

\bibitem{IETF_rfc9250}
Christian Huitema, Sara Dickinson, and Allison Mankin.
\newblock {DNS} over dedicated {QUIC} connections.
\newblock {\em {RFC}}, 9250:1--27, 2022.

\bibitem{DoQ_Vaibhav}
Mike Kosek, Luca Schumann, Robin Marx, Trinh~Viet Doan, and Vaibhav Bajpai.
\newblock {{DNS} Privacy with Speed?: Evaluating {DNS} over {QUIC} and its
  Impact on Web Performance}.
\newblock In {\em {IMC}}, pages 44--50. {ACM}, 2022.

\bibitem{WebPrivacy_Design_IFIP23}
Jayasree Sengupta, Mike Kosek, Justus Fries, Pratyush Dikshit, and Vaibhav
  Bajpai.
\newblock {Web Privacy By Design: Evaluating Cross-layer Interactions of QUIC,
  DNS and H/3}.
\newblock In {\em 2023 IFIP Networking Conference [\textit{To Appear}]}, pages
  1--9, 2023.

\bibitem{QUIC_CrossLayer_Jayasree}
Jayasree Sengupta, Mike Kosek, Justus Fries, Simone Ferlin, Pratyush Dikshit,
  and Vaibhav Bajpai.
\newblock On cross-layer interactions of quic, encrypted {DNS} and {HTTP/3:}
  design, evaluation and dataset.
\newblock {\em CoRR}, abs/2306.11643, 2023.

\bibitem{IETF_rfc9000}
Jana Iyengar and Martin Thomson.
\newblock {{QUIC:} {A} UDP-Based Multiplexed and Secure Transport}.
\newblock {\em {RFC}}, 9000:1--151, 2021.

\bibitem{mike:pam:2022}
Mike Kosek, Trinh~Viet Doan, Malte Granderath, and Vaibhav Bajpai.
\newblock {One to Rule Them All? {A} First Look at {DNS} over {QUIC}}.
\newblock In {\em Passive and Active Measurement Conference}, 2022.

\end{thebibliography}

\end{document}